\documentclass[11pt,a4paper]{article}
\usepackage{jheppub}

\usepackage{subcaption}
\usepackage{tikz-feynman}[luatex=true]
\usepackage{ctable}
\usepackage{graphicx,xcolor}
\usepackage{amsmath,amssymb,mathtools,slashed}

\usepackage{feynmp-auto}
\usepackage{simpler-wick}
\usepackage{kantlipsum}
\usepackage{float}
\usepackage{rotating}

\usepackage[T1]{fontenc} 
\usepackage{etoolbox}
\usepackage{extarrows} 
\usepackage{graphicx}
\usepackage{bm}
\usepackage{wasysym}
\usepackage{verbatim}

\usepackage{empheq}
\usepackage{multirow}
 
\usepackage{balance}

\usepackage[colorlinks=true,urlcolor=blue,linkcolor=blue]{hyperref}
\usepackage[capitalise]{cleveref}
\usepackage{siunitx}
\usepackage{enumerate}

\usepackage{array}
\usepackage{tcolorbox}
\usepackage{fontawesome}

\renewcommand{\vec}[1]{{\mathbf{#1}}}
\newcommand{\bX}{{\bf{X}}}

\newcommand{\p}{\partial}

\newcommand{\bx}{{\bf{x}}}

\newcommand{\E}{\boldsymbol{e}}

\newcommand{\mH}{\mathcal{H}}

\newcommand{\an}{\quad \textmd{and} \quad }
\newcommand{\bs}{\boldsymbol{\sigma}}

\newcommand{\bea}{\begin{eqnarray}}
\newcommand{\eea}{\end{eqnarray}}

\newcommand{\bq}{{\bf{q}}}

\newcommand{\bk}{{\bf{k}}}

\usepackage{upgreek}

\newcommand{\mpl}{M_{\mbox{\tiny{Pl}}}}
\newcommand{\e}{{\bf{\textsf{e}}}}

\newcommand{\ev}[1]{\ensuremath{\left\langle #1 \right\rangle}}

\title{Weyl Fermion Creation by Cosmological Gravitational Wave Background at 1-loop}
\author[a]{Azadeh Maleknejad}
\author[b,c]{Joachim Kopp}
\affiliation[a]{Department of Physics, King’s College London, Strand, London WC2R 2LS, UK}
\affiliation[b]{Theoretical Physics Department, CERN, 1211 Geneva 23, Switzerland}
\affiliation[c]{PRISMA Cluster of Excellence \& Mainz Institute for Theoretical Physics, \\
             Johannes Gutenberg University, Staudingerweg 7, 55099 Mainz, Germany}
             
\emailAdd{azadeh.maleknejad@kcl.ac.uk}
\emailAdd{jkopp@cern.ch}

\abstract{
Weyl fermions of spin $\frac12$ minimally coupled to Einstein's gravity in 4 dimensions cannot be produced purely gravitationally in an expanding Universe at tree level. Surprisingly, as we showed in a recent letter \cite{Maleknejad:2024ybn}, this changes at gravitational 1-loop when cosmic perturbations, like a gravitational wave background, are present. Such a background introduces a new scale, thereby breaking the fermions' conformal invariance. This leads to a non-vanishing gravitational self-energy for Weyl fermions at 1-loop and induces their production. In this paper, we present an extended study of this new mechanism, explicitly computing this effect using the in--in formalism. We work in an expanding Universe in the radiation-dominated era as a fixed background. Gravitational wave-induced fermion production has rich phenomenological consequences. Notably, if Weyl fermions eventually acquire mass, and assuming realistic -- and potentially detectable -- gravitational wave backgrounds, the mechanism can explain the abundance of dark matter in the Universe. More generally, gravitational-wave induced freeze-in is a new purely gravitational mechanism for generating other feebly interacting fermions, e.g.\ right-handed neutrinos. We show that this loop level effect can dominate over the conventional -- tree-level -- gravitational production of superheavy fermions in a sizable part of the parameter space.
\href{https://github.com/koppj/GW-freeze-in/}{\faGithub}
}

\keywords{}

\preprint{\newline
KCL-PH-TH/2024-33, 
CERN-TH-2024-079,
MITP-24-054}

\begin{document}

\maketitle

\section{Introduction}
\label{Sec:intro}

Recently in \cite{Maleknejad:2024ybn}, we have identified a new mechanism by which massless fermions minimally coupled to Einstein's gravity can be generated purely gravitationally. The key observation is that at 1-loop level, cosmic perturbations  -- for instance in the form of stochastic gravitational waves -- naturally break the conformal symmetry of Weyl fermions in General Relativity by introducing a new scale. Cosmological correlators are usually generated at tree level with subleading loop contributions. This effect, in contrast, starts at loop level. This raises the question of whether such perturbations can be responsible for the production of dark matter (DM) in the early Universe. Surprisingly, as we have shown in \cite{Maleknejad:2024ybn}, the answer is Yes. This is particularly relevant because the early Universe can be expected to be permeated by stochastic GWs: they could originate for example from gauge fields in inflation \cite{Sorbo:2011rz, Maleknejad:2016qjz, Komatsu:2022nvu}, first-order phase transitions \cite{Witten:1984rs, Schwaller:2015tja, Caprini:2015zlo, RoperPol:2023bqa}, primordial magnetic fields \cite{Brandenburg:2021aln, RoperPol:2021xnd}, preheating and gauge preheating \cite{Adshead:2019igv, Figueroa:2022iho}, cosmic strings \cite{Hindmarsh:1994re, Auclair:2019wcv}, etc.

The intriguing possibility that only gravitational interactions may be needed to create DM in the Universe has been discussed for a long time \cite{Ford:1986sy, Chung:1998zb, Parker_Toms_2009, Kolb:2023ydq}. The conventional mechanism, though, requires very massive fields with $M \gtrsim \SI{e13}{GeV}$ \cite{Kolb:2017jvz, Ema:2019yrd}. Such superheavy particles can be produced by the expansion of the Universe itself, a mechanism called Cosmological Gravitational Particle Production (CGPP) \cite{Kolb:2017jvz, Ema:2019yrd}. Alternatively, graviton-mediated annihilation of SM particles or inflatons can produce dark matter in a very hot plasma with temperatures $T_\text{reh} \gtrsim \SI{e13}{GeV}$ \cite{Garny:2015sjg, Bernal:2018qlk, Clery:2021bwz}. Production of \emph{massless} spin-$\frac12$ fermions (Weyl fermions), in contrast, is not possible based just on the expansion of the Universe. This is because generating massless fermions requires first breaking their conformal symmetry via interactions, for instance with the SM or the inflaton \cite{Greene:1998nh, Adshead:2018oaa, Maleknejad:2019hdr, Maleknejad:2020yys, Maleknejad:2020pec, Zhang:2023xcd}. In the context of inflation and massless fields, chiral gravitational waves (GWs) can also produce Weyl fermions when parity is broken in Chern--Simons gravity \cite{Alexander:2004us} or by non-Abelian gauge fields \cite{Maleknejad:2016dci, Maleknejad:2014wsa, Adshead:2015jza, Caldwell:2017chz}. The physical concept underlying the latter mechanism is anomalous chiral fermion production by the (global) gravitational anomaly \cite{Alvarez-Gaume:1983ihn, Eguchi:1976db} in the SM and beyond.  In \cref{tab:context} we compare the new gravitational fermion production mechanism introduced in \cite{Maleknejad:2024ybn} -- GW-induced freeze-in -- to the other gravitational production mechanisms known in the literature.

\begin{table}
    \centering
    \begin{tabular}[c]{p{4.8cm}@{\ }p{3.8cm}@{\ \ }p{4.2cm}c}
        \toprule
        \bf Production Mechanism
            & \bf\centering Underlying Physics
            & \bf Conditions
            & \bf Ref. \\
        \hline
        Cosmological Gravitational \newline Particle Production
            & \centering Cosmic expansion 
            & super-massive fields 
            & \cite{Kolb:2017jvz,Ema:2019yrd} \\
        \hline 
        Graviton-Mediated \newline Annihilation (GMA) & 
            \centering
            \raisebox{-1.3cm}{\resizebox{0.9\linewidth}{!}{
            \begin{tikzpicture}
              \begin{feynman}
                \vertex (b) ;
                \vertex [above left=of b] (i1) {\(SM\)};
                \vertex [below left=of b] (i2) {\(SM\)};
                \vertex [right=of b] (c);
                \vertex [above right=of c] (f1) {\(\Psi\)};
                \vertex [below right=of c] (f2) {\(\Psi\)};
                \diagram* {
                  (i1) -- [fermion] (b) -- [fermion] (i2),
                  (b) -- [gluon, red, edge label=\(h_{ij}\)] (c),
                  (c) -- [fermion] (f1),
                  (c) -- [fermion] (f2),
                };
              \end{feynman}
              \useasboundingbox ([shift={(-5mm,-5mm)}] current bounding box.south west) rectangle ([shift={(5mm,5mm)}] current bounding box.north east);
            \end{tikzpicture}}}
            & Super-massive field \newline High temperature plasma & \cite{Garny:2015sjg, Bernal:2018qlk,Clery:2021bwz} \\
        \hline
        Gravitational Leptogenesis &
            \centering
            \raisebox{-0.95cm}{\resizebox{0.9\linewidth}{!}{
            \begin{tikzpicture}
              \begin{feynman}
                \vertex (a) {\( \nabla_{\mu}J^{\mu}_5\)};
                \vertex [right=of a] (b);
                \vertex [above right=of b] (c);
                \vertex [below right=of b] (d);
                \vertex [right=of d] (e) ;
                \vertex [right=of c] (f);
                \vertex [above right=of f] (g);
                \vertex [below right=of f] (h);
                \diagram* {
                  (a) -- [dashed] (b) -- [fermion] (c),
                  (c) -- [gluon, red] (f),
                  (d) -- [fermion] (b),
                  (d) -- [gluon, red] (e),
                };
                \draw[fermion] (c) -- (d);
              \end{feynman}
              \useasboundingbox ([shift={(-5mm,-5mm)}] current bounding box.south west) rectangle ([shift={(5mm,5mm)}] current bounding box.north east);
            \end{tikzpicture}}}
            & Chiral GWs \newline Chiral fermions & \cite{Alexander:2004us,Maleknejad:2016dci}\\
        \hline
        GW-Induced Freeze-In &
            \centering
            \raisebox{-0.3cm}{\resizebox{0.9\linewidth}{!}{
            \begin{tikzpicture}
              \begin{feynman}
                \vertex (a) {\( \Psi\)} ;
                \vertex [right=of a] (b);
                \vertex [right=of b] (c) ;
                \vertex [right=of c] (d) {\(\Psi\)};
                 \vertex (2,1) (e);
                      \diagram* {
                  (a) -- [fermion] (b) -- [fermion] (c) -- [fermion] (d),
                  (b) -- [gluon, red, half left] (c) ,
                         };
              \end{feynman}
              \useasboundingbox ([shift={(-5mm,-5mm)}] current bounding box.south west) rectangle ([shift={(5mm,5mm)}] current bounding box.north east);
            \end{tikzpicture}}}
        & GW background & \cite{Maleknejad:2024ybn} \\[0.2cm]
        \bottomrule
    \end{tabular}
    \caption{Comparison of the new gravitational fermion production mechanism discussed in the present work -- GW-induced freeze-in -- to other gravitational production mechanisms known in the literature. The curly lines (in red) in the Feynman diagrams represent gravitons, i.e. $h_{ij}$.  All of the mechanisms listed here rely only on minimal couplings to Einstein gravity. For the comparison of the parameter regions in which different mechanisms can explain the dark matter relic density, see \cref{fig:OmegaD}.}
    \label{tab:context}
\end{table}

This paper provides an extended study of the findings presented in \cite{Maleknejad:2024ybn}. In particular, we show in detail how we evaluate the gravitational 1-loop diagrams, and from there the energy density of Weyl fermions produced in the presence of a stochastic GW background. If these fermions later acquire a mass (or have a small, but initially negligible, mass from the start), they can play the role of the DM today. We will discuss this possibility in detail as well. To obtain analytic results, we will work with a simple phenomenological broken power-law parameterization for the GW spectrum  during the radiation era presented in \cref{Fig:Omega}. This model provides a good fit to the results of simulations in many scenarios, e.g.\ phase transitions  \cite{Caprini:2009yp} and primordial magnetic fields \cite{Caprini:2018mtu}. We expect that our result is generic, but accurately estimating this phenomenon for other scenarios of primordial GWs requires advanced modeling and simulations, which we leave for future work.

The outline of this paper is as follows: in \cref{Sec:fermion-FRW} we introduce our formalism and discuss the interaction of Weyl fermions with GWs in a cosmological background. In \cref{Sec:fermion-1-loop}, we then compute the evolution of the fermion energy density in a stochastic GW background at 1-loop using the In--In formalism. Turning these results into concrete predictions requires a phenomenological parameterization of the GW background, which we introduce in \cref{Sec:GW}. In \cref{Sec:analytical}, we develop an analytical estimate of the energy density and pressure of the produced fermions. The phenomenological consequences of our results will be the topic of \cref{Sec:relic}. We conclude in \cref{Sec:conclusion} with a summary and a discussion of future directions. Our Notations and Conventions are collected in \cref{Sec:notation}. The spin connections and the interaction actions are worked out in \cref{Sec:app-spin-connection}.

Some of the numerical and computer algebra codes we have developed for this paper are available from \url{https://github.com/koppj/GW-freeze-in/}.

\section{Weyl Fermions in an Expanding Universe}
\label{Sec:fermion-FRW}

Consider Weyl fermions of spin-$\frac12$ in the early Universe during the radiation era. Their action is
\begin{align}
    S_{\psi} = \int\!\text{d}^4x \, \mathcal{L}_\psi
             = \frac{i}{2} \int\!\text{d}^4x \, \sqrt{-g} \
               \big[ \boldsymbol{\uppsi}^\dag_L \bar{\bs}^\mu
                     \overset{\leftrightarrow}{\mathcal{D}}_\mu \boldsymbol{\uppsi}_L
               + \boldsymbol{\uppsi}^\dag_R \bs^\mu
                     \overset{\leftrightarrow}{\mathcal{D}}_\mu \boldsymbol{\uppsi}_R \big] ,
    \label{Eq:FullAction-F}
\end{align}
where $\boldsymbol{\uppsi}_{R,L}$ are the right-/left-handed Weyl fermion fields, respectively, and $\overset{\leftrightarrow}{\mathcal{D}}_{\mu} = (\overset{\rightarrow}{\mathcal{D}}_{\mu} - \overset{\leftarrow}{\mathcal{D}}_{\mu})$. The spinor covariant derivative is $\mathcal{D}_\mu=\nabla_{\mu}-\omega_{\mu}$ where $\omega_{\mu}$ is the spin-connection. Moreover,  $\bs^{\mu} \equiv (\boldsymbol{I}, \bs^{i})$ and  $\bar{\bs}^{\mu} \equiv (\boldsymbol{I}, -\bs^{i})$. From this point forward, we focus on right-handed Weyl fermions and eliminate the ``$R$'' subscript. The case of left-handed fermions can be treated in complete analogy. The energy-momentum tensor of right-handed fermions is given as
\begin{align}
    T_{\mu\nu} = -\frac{2}{\sqrt{-g}} \frac{\delta (\sqrt{-g} \mathcal{L}_{\psi})}{\delta g^{\mu\nu}}
               = -\frac{i}{2}  \boldsymbol\uppsi^\dag \bs_{(\mu}
                       \overset{\leftrightarrow}{\mathcal{D}}_{\nu)} \boldsymbol{\uppsi} +  g_{\mu\nu} ~ \mathcal{L}_{\psi},
\end{align}
where $\bs_\mu = g_{\mu\nu} \bs^\nu$, and the symmetrization of the indices is denoted as ``$(\cdot, \cdot)$'', i.e.\ $X_{(i}Y_{j)} \equiv X_i Y_j + X_j Y_i$. In the Friedmann-Lema\^itre--Robertson--Walker (FLRW) geometry, the effect of the expansion of the Universe can be absorbed by redefining the fermion field as (see \cref{Eq:rescale} in \cref{Sec:app-spin-connection})
\begin{align}
    \boldsymbol{\Psi} \equiv a^{3/2} \boldsymbol{\uppsi},
\end{align}
where $\boldsymbol{\Psi}$ is the canonically normalized field and $a$ is the scale factor of the FLRW metric. The fermion energy density is diluted as $a^{-4}$ with the expansion of the Universe as a consequence of the conformal symmetry of massless Weyl fermions. Below, we will show that cosmic perturbations break this conformal symmetry and source the production of Weyl fermions.

Here and in the following, we will express the time-dependence of the fields in terms of conformal time $\tau$, which is related to cosmic time $t$ via the scale factor, i.e.
\begin{align}
    d\tau= \frac{dt}{a(t)}.   
\end{align}
The Weyl fermions can be decomposed into Fourier modes as
\begin{align}
    \boldsymbol{\Psi}(\tau,\bx) =
        \int\! d^3k \, 
            \boldsymbol{\Psi}_{\bf{k}}(\tau) ~ e^{i{\bf{k}}.{\bf{x}}}
        \,,
    \label{Eq:Psi-Weyl}
\end{align}
with the operator $\boldsymbol{\Psi}_{\bf{k}}(\tau)$ given as
\begin{align}
    \boldsymbol{\Psi}_{\bf{k}}(\tau) =
        \Big[
            {\bf{U}}_{ \bf{k}}(\tau) \, \hat{b}_\bk
         +  {\bf{V}}_{-\bf{k}}(\tau) \, \hat{c}^\dag_{-\bk}
        \Big] \,.
\end{align}
Here, $\hat{b}_\bk$ denotes the particle annihilation operators and $\hat{c}^\dag_\bk$ stands for the antiparticle creation operators. The corresponding spinors are $\bf{U}_\bk(\tau)$ and $\bf{V}_\bk(\tau)$; they are related by CP symmetry as
\begin{align}
    {\bf{V}}_\bk(\tau) = i \bs_2 {\bf{U}}^*_{-\bk}(\tau) .
    \label{eq:CP}
\end{align}
For the left-handed Weyl field, the relation is ${\bf{V}}_{L,\bf{k}}(\tau) = -i \bs_2 {\bf{U}}^*_{L,-\bf{k}}(\tau)$.

\subsection{Free Weyl Fermions in an FLRW Universe}
\label{Sec:free-fermions}

For free Weyl fermions in an FLRW Universe, the Fourier modes $\boldsymbol{\Psi}^{(0)}_{\bk}(\tau)$ can be found explicitly by solving the free Dirac equation, using the Bunch--Davies vacuum as the initial condition. This gives
\begin{align}
    {\bf{U}}^{(0)}_\bk = \frac{e^{-i k \tau}}{\sqrt{2} (2\pi)^{\frac32}} {\bf{E}}^+_\bk
                                      \an  
    {\bf{V}}^{(0)}_\bk = \frac{e^{ i k \tau}}{\sqrt{2} (2\pi)^{\frac32}} {\bf{E}}^-_{-\bk},
    \label{Eq:U0-V0}
\end{align}
where $k \equiv |\bk|$ and $\bf{E}^\pm_\bk$ are the helicity eigenstates
\begin{align}
    {\bf{E}}^+_\bk = \frac{\tilde{k}_\mu \bs^{\mu}}
                        {\sqrt{2 k (k + k_3)}} \begin{pmatrix} 1 \\ 0 \end{pmatrix}
\an
    {\bf{E}}^-_\bk = \frac{\tilde{k}_\mu \bar{\bs}^{\mu}}
                        {\sqrt{2 k (k + k_3)}} \begin{pmatrix} 0 \\ 1 \end{pmatrix},
    \label{Eq:E-pm}
\end{align}
with $\tilde{k}_{\mu} \equiv (k,\bk)$. Given that helicity and chirality are equivalent for free, massless fermions, the zeroth-order solution for left-handed Weyl fermions, is similar, but with $\bf{E}^+_\bk$ replaced by $\bf{E}^-_\bk$. The field equation of free Weyl fermions is not time-dependent, i.e.\ no particle production occurs \cite{Kolb:2023ydq}.

\subsection{Interaction with a Stochastic Gravitational Wave Background}

The metric of an expanding FLRW Universe permeated by a GW background is
\begin{align}
    ds^2 = -dt^2 + a^2(t) \, \hat{g}_{ij} \, dx_i dx_j,
    \label{Eq:metric}
\end{align}
where
\begin{align}
   \hat{g}_{ij} = \Big( \delta_{ij} + h_{ij} + \frac12 h_{ik} h_{jk} + \dots \Big),
   \qquad
   \det\hat{g}  = 1.
\end{align}
Here, $h_{ij}$ describes the metric perturbation due to the GW, for which we choose the transverse--traceless gauge.  In this work, we only consider perturbations due to the GW background during the radiation era, leaving the inclusion of cosmological curvature perturbations and cosmic inflation for future work \cite{in-preparation-II}. Notice that we used the conventional notation in cosmology literature (e.g.\ Refs.~\cite{Weinberg:2008zzc, Maggiore:2018sht}) in defining the perturbed metric as $g_{\mu\nu} = \bar{g}_{\mu\nu} + h_{\mu\nu}$. The alternative notion used in gravity literature (e.g.\ Refs.~\cite{Donoghue:2017pgk, Ortin:2015hya}) is defined as $g_{\mu\nu}=\bar{g}_{\mu\nu}+\frac{1}{\mpl} \tilde{h}_{\mu\nu}$.

During the radiation era, $a$, $\tau$, and $H$ are related as
\begin{align}
    \mathcal{H} \equiv a H = \frac{1}{\tau}, \qquad
    a =\frac{\tau}{\tau_* z_*},
    \label{eq:rad-era-relations}
\end{align} 
in which $\tau_*$ is a pivot time and $z_*$ is the redshift at $\tau_*$.

The metric perturbation can be decomposed into Fourier modes as
\begin{align}
    h_{ij}(\tau,\vec{x}) = \sum_{s=\pm} \int\!d^3q \, \hat{h}_{s,\bq}(\tau) \, \E^s_{ij}(\hat{\vec{q}})
                           \, e^{i \bq.\bx} \,,
    \label{Eq:GW}
\end{align}
where the sum runs over the two circular polarization states of GWs and the $h_{s,\bq}(\tau)$ operator is
\begin{align}
    \hat{h}_{s,\bq}(\tau) = \Big[ \mathrm{h}_{s,\bq}(\tau) \, \hat{a}^s_{\bq}
                          + \mathrm{h}^{*}_{s,-\bq}(\tau) \, \hat{a}^{s\dag}_{-\bq} \Big],
\end{align}
The Fourier coefficients $h_{\pm\bq}$ give the amplitudes of the Fourier mode with momentum $\bq$ and polarization $\pm$, while $\hat{a}^s_\bq$ denotes the canonically normalized graviton annihilation operators and $\E^\pm_{ij}(\hat{\vec{q}})$ are the circular polarization tensors of helicity $\pm 2$. The sum runs over these circular GW polarization states. The polarization tensors satisfy the normalization condition $\E^{s*}_{ij}(\hat{\vec{q}}) \, \E^{s'}_{ij}(\hat{\vec{q}}) = 2 \delta^{ss'}$.\footnote{The explicit form of $\E^{\pm}_{ij}(\hat{\vec{q}})$ in spherical coordinates is $\E^s_{ij}(\hat{\bq}) = \sqrt{2} \, \hat\E^s_{i}(\hat{\bq}) \,\otimes\, \hat{\E}^s_{j}(\hat{\bq})$, where $\hat{\E}^\pm(\hat{\bq}) = \frac{1}{\sqrt{2}}(\hat{\theta} \pm i\hat{\phi})$ are circular polarization vectors, with $\hat\theta$ and $\hat\phi$ two orthogonal unit vectors in the plane perpendicular to $\hat{q}$ \cite{Weinberg:2008zzc}.} 

Although in this work we consider unpolarized GWs, working with circular polarization states ($h_{\pm}$) is more convenient here than using plus- and cross-polarization states ($h_{+,\times}$), even though the latter are more common in the GW literature.  We have worked out the spin connections and interaction actions for the metric, \cref{Eq:metric}, in \cref{Sec:app-spin-connection}. Here we report the final results. 

\begin{figure}
    \centering
    \includegraphics[width=0.5\columnwidth]{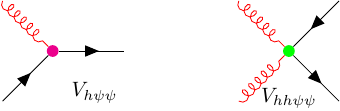}
    \caption{The graviton--fermion cubic and quartic vertices $V_{h\psi\psi}$ and $V_{hh\psi\psi}$ respectively, originated from $\mathcal{L}^{(1)}_\text{int}$ and $\mathcal{L}^{(2)}_\text{int}$.}
    \label{fig:vertices}
\end{figure}

\textbf{Cubic Interaction $V_{h\psi\psi}$:}
The first-order interaction Lagrangian between the Dirac field $\boldsymbol\Psi_{\!D} = (\boldsymbol\Psi_L, \boldsymbol\Psi_R)$ and GWs is (see \cref{Sec:1st-GW-fermion})
\begin{align}
    \mathcal{L}^{(1)}_\text{int} = -\frac{i}{2a^4} h_{ij} \bar{\boldsymbol{\Psi}}_{\!D} \gamma^{i}
                                    \overset{\leftrightarrow}{\p}_j \boldsymbol{\Psi}_{\!D}.
    \label{Eq:L1-int}
\end{align}
which corresponds to the cubic vertex $V_{h\psi\psi}$ in \cref{fig:vertices}. It is straightforward to find the associated interaction for the Weyl fermions from this.

\textbf{Quartic Interaction  $V_{hh\psi\psi}$:}
At second order in the metric perturbation, the fermion--GW interaction is (see \cref{Sec:2nd-GW-fermion})
\begin{align}
    \mathcal{L}^{(2)}_\text{int} = -\frac{i}{16a^3}  {\bf{e}}^{\mu}_{~\alpha} h_{ij} \p_{\mu} h_{ik}
                                    \bar{\boldsymbol{\Psi}}_{\!D} \Gamma^{\alpha j k} \boldsymbol{\Psi}_{\!D},
    \label{Eq:L2-int}
\end{align}
where $\Gamma^{\alpha j k}$ is the totally antisymmetrized product of three gamma matrices, and $ {\bf{e}}^\mu_{~\alpha}$ are the tetrads.

\section{Gravitational 1-Loop Corrections to Fermionic In--In Correlations}
\label{Sec:fermion-1-loop}

\begin{figure}[t]
    \centering
    \includegraphics[width=0.7\textwidth]{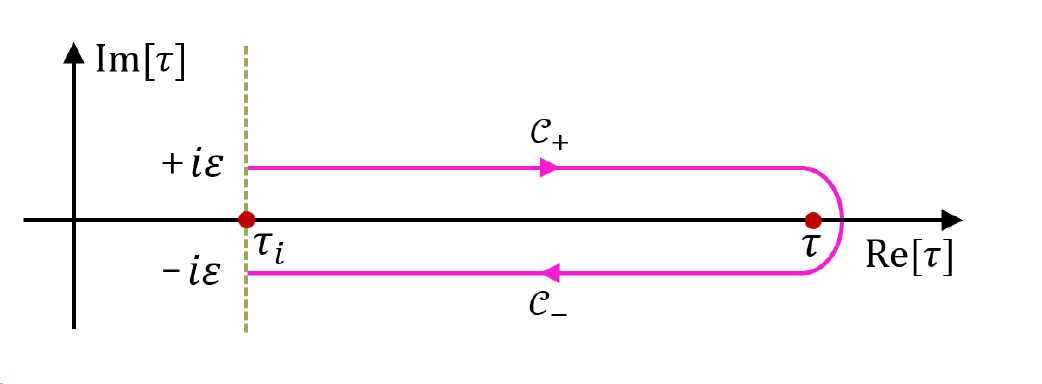}
    \caption{The Schwinger--Keldysh contour at zero temperature in the complex conformal time plane, $\tau$, starting and ending at $(1\pm i \epsilon)\tau_i$ were $\epsilon$ is an infinitesimal real quantity. $\mathcal{C}_+$ is the forward branch, and $\mathcal{C}_-$ is the backward branch.}
    \label{fig:SK}
\end{figure}

In this section, we compute the gravitational 1-loop corrections to the fermion two-point correlation function, taking into account the interactions $\mathcal{L}_\text{int}^{(1)}$ and $\mathcal{L}_\text{int}^{(2)}$.  We work in the In--In formalism, where the expectation value of an arbitrary operator $Q(t)$  \cite{Weinberg:2005vy} is
\begin{align}
    \big\langle Q(t) \big\rangle &=
        \bigg\langle \bar{\rm{T}} \exp\bigg[i \int_{t_i^-}^t dt''\, H_\text{int}(t'') \bigg]
                     Q_{I}(t) \,
                     {\rm{T}} \exp\bigg[-i \int_{t_i^+}^t dt' \, H_\text{int}(t') \bigg] \bigg\rangle\,,
\end{align}
where ($\bar{\rm{T}}$) $\rm T$ denotes (anti-)time ordering, $Q_I(t)$ is the operator $Q(t)$ in the interaction picture, and $t_i^{\pm} \equiv  (1 \pm i \epsilon)t_i$ is a time variable with an infinitesimal imaginary part (see \cref{fig:SK}). The interaction Hamiltonian is
\begin{align}
    H_\text{int}(t) = -\int \! d^3x \, a^3(t) \, \mathcal{L}_\text{int}(t,\bx) ,
\end{align}
where $\mathcal{L}_\text{int} = \mathcal{L}^{(1)}_\text{int} + \mathcal{L}^{(2)}_\text{int}$.

The quantity of our interest is the energy density of Weyl fermions,
\begin{align}
    \rho_\psi(\tau,\bx) = T_{\mu\nu} n^{\mu} n^{\nu}
                        = \frac{i}{a^4} \boldsymbol{\Psi}^{\dag}
                          \overset{\leftrightarrow}{\p_{\tau}} \boldsymbol{\Psi} - \mathcal{L}_{\psi},
    \label{Eq:rho-Interaction}          
\end{align}
where $n^{\mu}$ is the unit normal to equal time hypersurfaces. We have used the fact that in the presence of GWs, $n^\mu = (1,0,0,0)$. Note that $\mathcal{L}_\psi$ vanishes on-shell but must be considered for off-shell fermions. In the following, we compute the contribution of $\mathcal{L}^{(1)}_\text{int}$ and $\mathcal{L}^{(2)}_\text{int}$ to the fermion energy density. We note that in the context of chiral gravitational waves (GWs), a topic beyond the scope of the present study, the triangle diagram in \cref{tab:context} corresponding to the global gravitational anomaly should also be considered \cite{Alvarez-Gaume:1983ihn}. See for instance Refs.~\cite{Alexander:2004us}  and \cite{Maleknejad:2016dci}.

\subsection{Contribution of the Cubic Vertex}

We start by computing the contribution of the first-order interaction (cubic vertex) $\mathcal{L}^{(1)}_\text{int}$ to the energy density. Expanding the In--In expectation value $\ev{\rho_\psi(\tau,\bx)}\!_{\mathcal{L}^{(1)}_\text{int}}$ up to second order in $\mathcal{L}^{(1)}_\text{int}$, we obtain
\begin{align}
    \ev{\rho_\psi(\tau, \bx)}\!_{\mathcal{L}^{(1)}_\text{int}} \equiv \varrho_1 + \varrho_2,
\end{align}
where
\begin{align}
    \varrho_1 &\equiv \Big\langle \Big( \int \! d^3x'' \! \int_{\tau^-_i}^\tau \! d\tau''
                      a^4(\tau'') \, \mathcal{L}^{(1)}_\text{int}(\tau'', \bx'') \Big) \,
                      \rho^{(0)}_\psi(\tau,\bx) \,
                      \Big( \int \! d^3x' \! \int_{\tau^{+}_i}^\tau \! d\tau'
                      a^4(\tau') \, \mathcal{L}^{(1)}_\text{int}(\tau', \bx') \Big) \Big\rangle \nonumber\\
              &-\text{Re} \Big[ \Big\langle \bar{\rm{T}} \Big(
                                    \int \! d^3x'' \, d^3x' \int_{\tau_i^-}^\tau \! d\tau'' \, d\tau' \,
                                    a^4(\tau'') \, a^4(\tau') \, \mathcal{L}^{(1)}_\text{int}(\tau'',\bx'') \,
                                    \mathcal{L}^{(1)}_\text{int}(\tau',\bx') \Big) \,
                                    \rho^{(0)}_\psi(\tau,\bx)
                                \Big\rangle \Big],
    \label{Eq:rho-1}
\intertext{and}
    \varrho_2 &\equiv \Big\langle \Big( -i\int \! d^3x' \int_{\tau^-_i}^\tau \! d\tau' \,
                          a^4(\tau') \, \mathcal{L}^{(1)}_\text{int}(\tau') \Big) \, \rho^{(1)}_\psi(\tau,\bx)
                      \Big\rangle + c.c..
    \label{Eq:rho-2}
\end{align}
Diagrammatically, $\varrho_1$ and $\varrho_2$ can be represented as
\begin{align}
	\varrho_1 = \ev{ \rho_{\psi}^{(0)}	\mathord{
		\begin{tikzpicture}[radius=1.5cm, scale=0.6, baseline=-0.65ex, very thick]
 		      \draw[decoration={markings, mark=at position 0.3 with {\arrow[scale=-1.5, rotate=6]{stealth}},
                  mark=at position 0.6  with {\arrow[scale=-1.5, rotate=6]{stealth}},
                  mark=at position 0.98 with {\arrow[scale=-1.5, rotate=7]{stealth}}},
                  postaction=decorate](0, 0) circle [];
            \draw (-1.5,0) node {\huge $\times$}; 
             \begin{feynman}
				\vertex (A) at (1.125,0.975);
				\vertex (C) at (0,0);
				\coordinate (D) at (-1.5, 0);
				\vertex (B) at (1.125, -0.975); 
				\diagram*{
					(A) -- [gluon, red, bend right] (B)
			    };
       \filldraw[white] (0.8,0) circle [radius=0.22cm];
          \draw[red, very thick] (0.8,0) node {\large{$\boldsymbol{\otimes}$}}; 
         \end{feynman} 
	    \end{tikzpicture} } } ,
    \qquad\qquad
    \varrho_2 = \ev{ \rho_{\psi}^{(1)}	\mathord{
		\begin{tikzpicture}[radius=1.5cm, scale=0.6, baseline=-0.65ex, very thick]
 		    \draw[decoration={markings, mark=at position 0.3 with {\arrow[scale=-1.5, rotate=6]{stealth}},
                  mark=at position 0.6  with {\arrow[scale=-1.5, rotate=6]{stealth}},
                  mark=at position 0.98 with {\arrow[scale=-1.5, rotate=7]{stealth}}},
                  postaction=decorate] (0, 0) circle [];
            \draw (-1.5,0) node {\huge $\times$}; 
  			\begin{feynman}
				\vertex (A) at (1.125,0.975);
				\vertex (C) at (0,0);
				\coordinate (D) at (-1.5, 0);
				\vertex (B) at (1.125, -0.975); 
				\diagram*{
					(A) -- [gluon, red] (D)
			    };
          \filldraw[white] (-0.18,0.54) circle [radius=0.22cm];
          \draw[red, very thick] (-0.18,0.54) node {\large{$\boldsymbol{\otimes}$}}; 
   			\end{feynman} 
        \end{tikzpicture} } },
	\label{Eq:loop}
\end{align} 
where the curly lines (in red) represent background GWs, solid lines (in black) are fermions, $\otimes$ denotes the background gravitational wave bath in the early Universe, and $\times$ stands for the insertion of the operator $\rho_\psi^{(0)}(\tau,\bx)$ or $\rho_\psi^{(1)}(\tau,\bx)$. In the following, we evaluate these two diagrams.

\subsubsection*{The $\varrho_1$ term}
\label{Sec:rho-1}

\Cref{Eq:rho-1} can be decomposed as
\begin{align}
    \varrho_1(\tau) \equiv \frac{1}{a^4(\tau)} \Big[ \mathcal{G}_1(\tau) + \mathcal{G}_2(\tau)\Big],
\end{align}
where up to a factor $a^4(\tau)$, $\mathcal{G}_1$ is the first line of \cref{Eq:rho-1}, i.e.
\begin{align}
     \mathcal{G}_1(\tau) \equiv
     \Big\langle \! \bigg( \! \int\!\!d^3x'' \!\!\!
                             \int_{\tau^-_i}^\tau \!\!\! d\tau''
        a^4(\tau'') \mathcal{L}^{(1)}_\text{int}(\tau'',\bx'') \!\bigg)
        a^4(\tau) \rho^{(0)}_\psi(\tau,\bx) \,
        \bigg(\!\int \!\! d^3x' \!\! \int_{\tau^{+}_i}^\tau \!\! d\tau'
        a^4(\tau') \mathcal{L}^{(1)}_\text{int}(\tau',\bx') \!\bigg) \!
    \Big\rangle
    \label{Eq:cG1}
\end{align}
and $\mathcal{G}_2$ is similarly related to the second line of \cref{Eq:rho-1}:
\begin{align}
    \mathcal{G}_2(\tau) \!\equiv\!
    -\text{Re} \Big[ \Big\langle \bar{\rm{T}} \!
        \int \! d^3x'' \, d^3x' \!\!
        \int_{\tau_i^-}^\tau \!\! d\tau'' d\tau'
        a^4(\tau'') \, a^4(\tau') \mathcal{L}^{(1)}_\text{int}(\tau'',\bx'')
         \mathcal{L}^{(1)}_\text{int}(\tau',\bx') \, a^4(\tau)
        \rho^{(0)}_\psi(\tau,\bx)
    \Big\rangle \Big]
    \label{Eq:cG2}
\end{align}
In the following, we begin by calculating $\mathcal{G}_1$, followed by the computation of $\mathcal{G}_2$.

\begin{figure}
    \centering
    \includegraphics[scale=0.4]{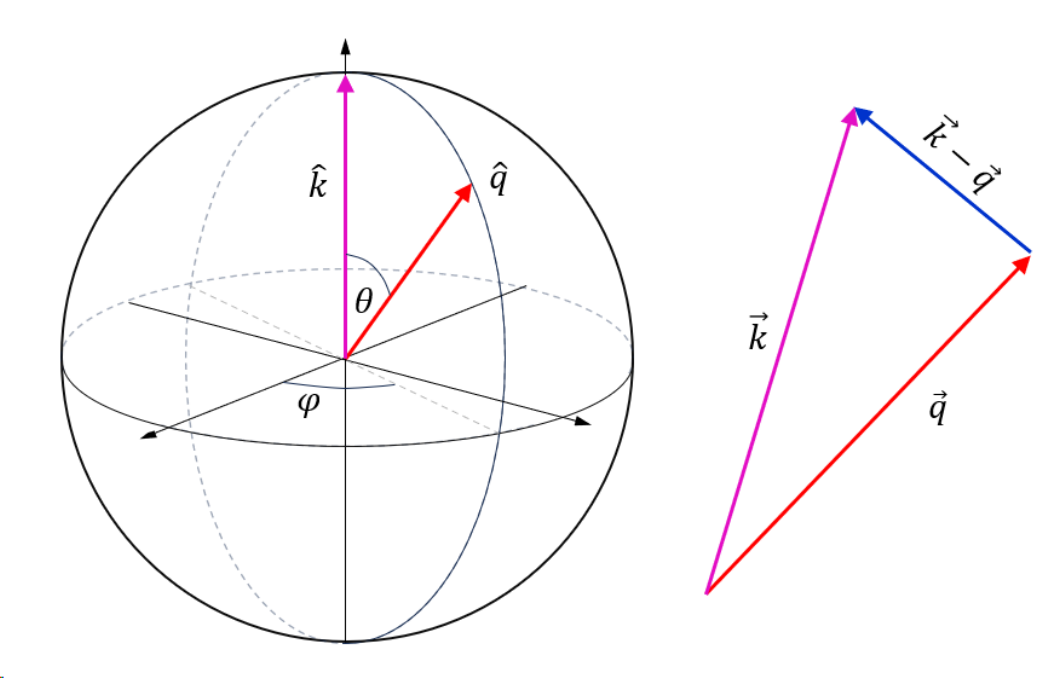}
    \caption{The coordinate system used in this work, with $\bk = k\hat{z}$ aligned along the $z$-axis, and $\bq$ at an angle $(\theta,\phi)$. We use $\omega$ as the frequency of $\bk-\bq$, i.e. $\omega = \sqrt{k^2+q^2-2k q \cos\theta}$.}
    \label{fig:angular}
\end{figure}   


\vspace{1ex}
\noindent
\textbf{\textit{i) The $\mathcal{G}_1$ term:}}
The first step in evaluating \cref{Eq:cG1} is to expand the fermion field operators as well as the metric perturbation into Fourier modes and to factor the expectation value, $\ev{\,\cdot\,}$, into an expectation value over the fermionic current and an expectation value over the GW degrees of freedom. Assuming stochastically isotropic and circularly unpolarized GWs, the latter contribution can be simplified as
\begin{align}
    \ev{\hat{h}_{s,\bq'}(\tau'') \, \hat{h}\dag_{s,\bq}(\tau')}
        = \big\langle \mathrm{h}_q(\tau'') \, \mathrm{h}^*_q(\tau') \big\rangle
          \delta_{ss'} \delta^{(3)}(\bq-\bq').
    \label{Eq:power-GW}
\end{align}
The resulting $\delta$-function can be used to eliminate one of the eight momentum integrals that appear from the Fourier decomposition of the fields. Two more momentum-conserving $\delta$-functions arise from the integrals over $\bx'$ and $\bx''$. The remaining expression is
\begin{multline}
    \mathcal{G}_1(\tau) = (2\pi)^6 \, i \int_{\tau^+_i}^\tau \! d\tau' \! \int_{\tau^-_i}^\tau \! d\tau''
                          \int_\bq \! \big\langle \mathrm{h}_q(\tau'') \, \mathrm{h}^*_q(\tau') \big\rangle 
                          \int_{\bk''} \int_{\bk'} k' k'' \sum_{s=\pm} \int_{\bk_1} \int_{\bk_2}
                          e^{i(\bk_1-\bk_2).\bx} \\
                          \times \Big\langle
                              \boldsymbol{\Psi}^{\dag}_{\bk''}(\tau'') \,
                              \bX^{s}_{\hat\bk'',\hat\bq} \,
                              \boldsymbol{\Psi}_{\bk''-\bq}(\tau'') \,
                              \boldsymbol{\Psi}^{\dag}_{\bk_2}(\tau) \,
                              \overset{\leftrightarrow}{\p}_{\tau}
                              \boldsymbol{\Psi}_{\bk_1}(\tau) \,
                              \boldsymbol{\Psi}^{\dag}_{\bk'-\bq}(\tau') \,
                              \bX^{s\dag}_{\hat\bk',\hat\bq} \,
                              \boldsymbol{\Psi}_{\bk'}(\tau')
                          \Big\rangle,
    \label{Eq:cG1-B}
\end{multline}
where we have used the shorthand notation $\int_\bk \equiv \int d^3k$. The fermion fields $\boldsymbol\Psi_\bk(\tau)$ in \cref{Eq:cG1-B} are meant to be the \emph{free} fields $\Psi^{(0)}_\bk(\tau)$ from \cref{Sec:free-fermions}, but to avoid cluttering we have omitted the superscript $(0)$. For the interaction vertex, we have introduced the shorthand notation
\begin{align}\label{Eq:Xs}
    \boldsymbol{X}^s_{\hat{\bk},\hat{\bq}} \equiv
        {\boldsymbol{\sigma}}.\boldsymbol{e}^s(\hat{\bq}).\hat{\bk},   
\end{align}
where $\hat\bk$ is a unit vector in the direction of $\bk$.

The fermion 6-point function in \cref{Eq:cG1-B} can be contracted as
\begin{align}
    \langle ...\rangle &= \Big\langle\wick{
        \color{red}     \c1{\boldsymbol\Psi}^\dag_{\bk''}(\tau'') \,
        \color{black}   \bX^{s}_{\hat\bk'',\hat\bq}
        \color{olive}   \c2{\boldsymbol\Psi}_{\bk''-\bq}(\tau'') \,
        \color{magenta} \c3{\boldsymbol\Psi}^\dag_{\bk_2}(\tau)
        \color{black}   \overset{\leftrightarrow}{\p}_{\tau} 
        \color{red}     \c1{\boldsymbol\Psi}_{\bk_1}(\tau) \,
        \color{olive}   \c2{\boldsymbol\Psi}^{\dag }_{\bk'-\bq}(\tau') \,
        \color{black}   \bX^{s\dag}_{\hat\bk',\hat\bq}
        \color{magenta} \c3{\boldsymbol\Psi}_{\bk'}(\tau')
    } \Big\rangle  \nonumber\\
    &+ \Big\langle\wick{
        \color{magenta} \c1{\boldsymbol\Psi}^\dag_{\bk''}(\tau'') \,
        \color{black}   \bX^{s}_{\hat\bk'',\hat\bq}
        \color{olive}   \c2{\boldsymbol\Psi}_{\bk''-\bq}(\tau'') \,
        \color{olive}   \c2{\boldsymbol\Psi}^\dag_{\bk_2}(\tau)
        \color{black}   \overset{\leftrightarrow}{\p}_{\tau}
        \color{red}     \c2{\boldsymbol\Psi}_{\bk_1}(\tau) \,
        \color{red}     \c2{\boldsymbol\Psi}^\dag_{\bk'-\bq}(\tau') \,
        \color{black}   \bX^{s\dag}_{\hat\bk',\hat\bq}
        \color{magenta} \c1{\boldsymbol\Psi}_{\bk'}(\tau')
    } \Big\rangle  \nonumber\\
    &+ \Big\langle\wick{
        \color{magenta} \c1{\boldsymbol\Psi}^\dag_{\bk''}(\tau'') \,
        \color{black}   \bX^{s}_{\hat\bk'',\hat\bq}
        \color{olive}   \c2{\boldsymbol\Psi}_{\bk''-\bq}(\tau'') \,
        \color{red}     \c3{\boldsymbol\Psi}^\dag_{\bk_2}(\tau)
        \color{black}   \overset{\leftrightarrow}{\p}_\tau
        \color{red}     \c3{\boldsymbol\Psi}_{\bk_1}(\tau) \,
        \color{olive}   \c2{\boldsymbol\Psi}^\dag_{\bk'-\bq}(\tau') \,
        \color{black}   \bX^{s\dag}_{\hat\bk',\hat\bq} 
        \color{magenta} \c1{\boldsymbol\Psi}_{\bk'}(\tau')
    } \Big\rangle.
    \label{Eq:contractions-G1}
\end{align}
Here, we have used different colors in addition to the usual brackets to highlight the contractions. The last line describes a disconnected (vacuum) graph, which does not contribute to physical observables.

We now rewrite the vertices $\bX^s_{\hat\bk,\hat\bq}$ as
\begin{align}
    \boldsymbol{X}^s_{\hat\bk,\hat\bq} \, {\bf{E}}^\lambda_{\bf{k-q}}
        = \sum_{\lambda'=\pm} f^{s\lambda}_{\lambda'}(\hat{\bq},\hat{\bk}) \,
          {\bf{E}}^{\lambda'}_\bk ,
    \label{Eq:XE-dec}
\end{align}
where ${\bf{E}}^\lambda_\bk$ are the helicity eigenstates from \cref{Eq:E-pm}, and $f^{s\lambda}_{\lambda'}(\hat{\bq},\hat{\bk})$ are eight functions, with $s = \pm 2$ labeling the GW helicity, $\lambda,\lambda' = \pm\frac12$ denoting the helicity of the fermions. Without loss of generality we can consider $\bk$ in the $\hat{z}$ direction and find
\begin{align}
    \boldsymbol{X}^+_{\hat{\bk},\hat{\bq}} \, {\bf{E}}^{+}_{{\bf{k-q}}}
        &= \frac{(\omega+k+q) \sin\theta}{2\sqrt{\omega(\omega+k-q\cos\theta)}}
           \begin{pmatrix}
               \sin\theta \\
               e^{i\phi} (1-\cos\theta)
           \end{pmatrix} , \\
    \boldsymbol{X}^-_{\hat{\bk},\hat{\bq}} \, {\bf{E}}^{+}_{{\bf{k-q}}}
        &= \frac{(\omega+k-q) \sin\theta}{2\sqrt{\omega(\omega+k-q\cos\theta)}}
           \begin{pmatrix}
               \sin\theta \\
               -e^{i\phi} (1+\cos\theta) 
           \end{pmatrix}, \\
    \boldsymbol{X}^+_{\hat{\bk},\hat{\bq}} \, {\bf{E}}^{-}_{\bf{k-q}}
        &= -\frac{(\omega+k-q) \sin\theta}{2\sqrt{\omega(\omega+k-q\cos\theta)}}
           \begin{pmatrix}
               e^{-i\phi} (1+\cos\theta) \\
               \sin\theta
           \end{pmatrix}, \\
    \boldsymbol{X}^-_{\hat{\bk},\hat{\bq}} \, {\bf{E}}^{-}_{{\bf{k-q}}}
        &= \frac{(\omega+k+q) \sin\theta}{2\sqrt{\omega(\omega+k-q\cos\theta)}}
           \begin{pmatrix}
               e^{-i\phi} (1-\cos\theta) \\
               -\sin\theta
           \end{pmatrix},
\end{align}
where $\omega = \sqrt{k^2+q^2-2kq\cos\theta}$, and $\theta$ is the angle between $\bk$ and $\bq$ (see \cref{fig:angular}). Using \cref{eq:CP}, in evaluating the fermion 6-point function in \cref{Eq:contractions-G1}, we encounter expressions of the form
\begin{align}
    \mathcal{D}^\lambda(\hat\bq,\hat\bk) \equiv
        \sum_s  
               \big( \boldsymbol{E}^{-\lambda\dag}_\bk
                     \bX^s_{\hat\bk,\hat\bq}
                     \boldsymbol{E}^{\lambda}_{\bk-\bq}
               \big) \, \big( \boldsymbol{E}^{\lambda\dag}_{\bk-\bq}
                     \bX^{s\dag}_{\hat\bk,\hat\bq}
                     \boldsymbol{E}^{-\lambda}_\bk \big)
      = \sum_s \Big\lvert\big( \boldsymbol{E}^{-\lambda\dag}_\bk \bX^s_{\hat\bk,\hat\bq}
                               \boldsymbol{E}^{\lambda}_{\bk-\bq} \big) \Big\lvert^2,    
\end{align}
where $\lambda = \pm$ distinguishes between right-handed and  left-handed fermions. We find that
\begin{align}
    \mathcal{D}^+(\hat\bq,\hat\bk) = \mathcal{D}^-(\hat\bq,\hat\bk)
        = 2\sin^2\theta - \frac12 \frac{(\omega+k)^2 + q^2}{\omega(\omega + k - q\cos\theta)} \, \sin^4\theta
        \equiv \mathcal{D}(\hat\bq,\hat\bk),
    \label{Eq:C-qk}
\end{align}
which is a positive quantity vanishing around $\theta=0$ and $\pi$, and has a maximum around $\theta\approx \frac{3\pi}{5}$.
Note that $\boldsymbol{X}^s_{\hat\bk,\hat\bq} \propto \sin\theta$, and the cubic interaction therefore vanishes for fermions propagating parallel to the propagation direction of GWs, i.e., $\bk \parallel \bq$. 

With the help of \cref{Eq:C-qk}, the fermion 6-point function in \cref{Eq:contractions-G1} can now be evaluated (keeping in mind that fermion fields anti-commute):
\begin{align}
    (2\pi)^6 \, i \int_{\bk_1} \int_{\bk_2} \int_{\bk''} k' k'' \, e^{i (\bk_1-\bk_2).\bx} \langle \dots \rangle
  = \frac{1}{4 (2\pi)^3} (k'+\omega') k'^2 \, e^{i(k'+\omega')(\tau'-\tau'')} \,
    \mathcal{D}(\hat\bq,\hat\bk') + c.c. \,,
\end{align}
where, $\omega' \equiv \sqrt{k'^2+q^2-2k'q\cos\theta}$. For $\mathcal{G}_1(\tau)$, this leads to
\begin{align}
    \mathcal{G}_1(\tau) &= \frac{1}{4\pi} \int_{\tau^+_i}^{\tau} d\tau' \int_{\tau^-_i}^{\tau} d\tau''
        \int \! q^2 dq \, \big\langle \mathrm{h}_{q}(\tau'') \mathrm{h}^*_{q}(\tau') \big\rangle
        \int \! k^4 dk \int \! d\theta \, \sin^3\theta \, e^{i(k+\omega)(\tau'-\tau'')}
                                                                                \nonumber\\
    &\times (k+\omega) \Big[ 2 - \frac{\sin^2\theta((\omega+k)^2+q^2)}{2\omega(\omega+k-q\cos\theta)}\Big] + c.c. \,.
    \label{Eq:cG1-C}
\end{align}


\vspace{1ex}
\noindent
{\textbf{\textit{ii) The $\mathcal{G}_2$ term:}}}
We now evaluate \cref{Eq:cG2} in the same way as we have evaluated $\mathcal{G}_1$. We begin again by exploiting the isotropy and lack of polarization of the GW background, and we expand the fields in Fourier modes. We find
\begin{multline}
    \mathcal{G}_2(\tau) =
        (2\pi)^6 \,{\rm Im} \bigg[ i \int_{\tau^+_i}^{\tau} d\tau' \int_{\tau^-_i}^{\tau} d\tau''
        \int_\bq \! \big\langle \mathrm{h}_q(\tau'') \, \mathrm{h}^*_q(\tau') \big\rangle 
        \int_{\bk''} \int_{\bk'} k' k'' \sum_{s=\pm} \int_{\bk_1} \int_{\bk_2} e^{i(\bk_1-\bk_2).\bx} \\
    \times \Big\langle
        \boldsymbol\Psi^\dag_{\bk''}(\tau'') \,
        \bX^{s\dag}_{\hat{\bk''},\hat{\bq}} \,
        \boldsymbol\Psi_{\bk''-\bq}(\tau'') \,
        \boldsymbol\Psi^\dag_{\bk'-\bq}(\tau') \,
        \bX^{s}_{\hat{\bk'},\hat{\bq}}
        \boldsymbol\Psi_{\bk'}(\tau') \,
        \boldsymbol\Psi^\dag_{\bk_2}(\tau)
        \overset{\leftrightarrow}{\p}_{\tau}
        \boldsymbol\Psi_{\bk_1}(\tau)
    \Big\rangle \bigg].
    \label{Eq:cG2-}
\end{multline}
The fermion 6-point function can be contracted as
\begin{align}
    \langle ...\rangle &= \Big\langle\wick{
        \color{red}     \c1{\boldsymbol\Psi}^\dag_{\bk''}(\tau'') \,
        \color{black}   \bX^{s\dag}_{\hat{\bk''},\hat{\bq}}
        \color{olive}   \c2{\boldsymbol\Psi}_{\bk''-\bq}(\tau'') \,
        \color{olive}   \c2{\boldsymbol\Psi}^\dag_{\bk'-\bq}(\tau') \,
        \color{black}   \bX^{s}_{\hat{\bk'},\hat{\bq}} 
        \color{magenta} \c2{\boldsymbol\Psi}_{\bk'}(\tau') \,
        \color{magenta} \c2{\boldsymbol\Psi}^\dag_{\bk_2}(\tau) \,
        \color{black}   \overset{\leftrightarrow}{\p}_\tau
        \color{red}     \c1{\boldsymbol\Psi}_{\bk_1}(\tau)
    } \Big\rangle  \nonumber\\
    &+ \Big\langle\wick{
        \color{magenta} \c1{\boldsymbol\Psi}^\dag_{\bk''}(\tau'') \,
        \color{black}   \bX^{s\dag}_{\hat{\bk''},\hat{\bq}} 
        \color{olive}   \c2{\boldsymbol\Psi}_{\bk''-\bq}(\tau'') \,
        \color{red}     \c3{\boldsymbol\Psi}^\dag_{\bk'-\bq}(\tau') \,
        \color{black}   \bX^{s}_{\hat{\bk'},\hat{\bq}} 
        \color{magenta} \c1{\boldsymbol\Psi}_{\bk'}(\tau') \,
        \color{olive}   \c2{\boldsymbol\Psi}^\dag_{\bk_2}(\tau) \,
        \color{black}   \overset{\leftrightarrow}{\p}_{\tau}
        \color{red}     \c3{\boldsymbol\Psi}_{\bk_1}(\tau)
    } \Big\rangle \nonumber\\
    &+ \Big\langle\wick{
        \color{magenta} \c1{\boldsymbol\Psi}^\dag_{\bk''}(\tau'') \,
        \color{black}   \bX^{s\dag}_{\hat{\bk''},\hat{\bq}}
        \color{olive}   \c2{\boldsymbol\Psi}_{\bk''-\bq}(\tau'') \,
        \color{olive}   \c2{\boldsymbol\Psi}^\dag_{\bk'-\bq}(\tau') \,
        \color{black}   \bX^{s}_{\hat{\bk'},\hat{\bq}}
        \color{magenta} \c1{\boldsymbol\Psi}_{\bk'}(\tau') \,
        \color{red}     \c2{\boldsymbol\Psi}^\dag_{\bk_2}(\tau) \,
        \color{black}   \overset{\leftrightarrow}{\p}_{\tau}
        \color{red}     \c2{\boldsymbol\Psi}_{\bk_1}(\tau)
    } \Big\rangle .
    \label{Eq:contractions-G2}
\end{align}
The last line is once again a disconnected graph, which vanishes. Moreover, the first and second lines also vanish since they are both proportional to
\begin{align}  
\boldsymbol{\Psi}^\dag_{\bk_2}(\tau)  \overset{\leftrightarrow}{\p}_{\tau}
      \boldsymbol{\Psi}_{\bk_1}(\tau) \delta^{(3)}(\bk_1-\bk_2)\propto \boldsymbol{E}^{\lambda\dag}_{\bk_1} \boldsymbol{E}^{-\lambda}_{\bk_1} =0.
\end{align}      
As a result, the fermion six-point function in \cref{Eq:cG2-} vanishes, hence
\begin{align}
 \mathcal{G}_2(\tau) = 0,
\end{align}
and $\rho_1$ is simply given by $\mathcal{G}_1$ from \cref{Eq:cG1-C}:
\begin{align}
    \varrho_1(\tau) &= \frac{1}{4\pi} \frac{1}{a^4(\tau)}
        \int_{\tau^+_i}^{\tau} d\tau' \int_{\tau^-_i}^{\tau} d\tau''
        \int \! q^2 dq \, \big\langle \mathrm{h}_{q}(\tau'') \mathrm{h}^*_{q}(\tau') \big\rangle
        \int \! k^4 dk \int \! d\theta \, \sin^3\theta \, e^{i(k+\omega)(\tau'-\tau'')}
                                                                \nonumber\\
    &\times  (k+\omega) \Big[ 2 - \frac{\sin^2\theta((\omega+k)^2+q^2)}{2\omega(\omega+k-q\cos\theta)}\Big] + c.c. \,.
    \label{Eq:varrho-1-}
\end{align}
Let us pause here and take a closer look at the asymptotic form of the time integrals. In this work, we are interested in GW spectra that evolve dynamically between an initial time, $\tau_i$, and a ``settling time'', $\tau_*$. After the time $\tau_*$ where GW production ends and the GW background has reached a steady state, the time integrals approach a constant in time, i.e.
\begin{align}
    \lim_{\tau\gg \tau_*} \int_{\tau^+_i}^{\tau} d\tau' \int_{\tau^-_i}^{\tau} d\tau'' \,
    \big\langle \mathrm{h}_{q}(\tau'') \mathrm{h}^*_{q}(\tau') \big\rangle \,
    e^{i(k+\omega)(\tau'-\tau'')} \equiv f(k,q,\omega).
\end{align}
For explicit computations of these integrals in a phenomenological model for the GW background, see \cref{eq:T-incoh,eq:T-coh}. As a result, the asymptotic form of \cref{Eq:varrho-1-} for $\tau \gg\tau_*$ is $\varrho_1(\tau) \propto 1 / a^4(\tau)$.

\subsubsection*{The $\varrho_2$ term}

Expanding the fields appearing in \cref{Eq:rho-2} in Fourier space, we can write $\varrho_2$ as
\begin{align}
    \varrho_2(\tau) &= -\frac{4 (2\pi)^3}{a^4(\tau)} \int d^3q \int d^3k \, k^2 \,
        \text{Im}\Big[ \int_{\tau^{+}_i}^\tau d\tau' \,
        \big\langle \mathrm{h}_{q}(\tau) \, \mathrm{h}^*_q(\tau') \big\rangle
                                         \nonumber\\
    &\times \sum_s \Big\langle
        \boldsymbol\Psi^\dag_{\bk-\bq}(\tau) \,
        \boldsymbol{X}^{s\dag}_{\hat\bk,\hat\bq}
        \boldsymbol\Psi_\bk(\tau) \,
        \boldsymbol\Psi^\dag_\bk(\tau') \,
        \boldsymbol{X}^s_{\hat\bk,\hat\bq}
        \boldsymbol\Psi_{\bk-\bq}(\tau') 
    \Big\rangle \Big].
\end{align}
The fermion 4-point function in the second line is
\begin{align}
     \langle \dots \rangle &= \Big\langle\wick{
         \color{red}   \c1{\boldsymbol\Psi}^\dag_{\bk-\bq}(\tau) \,
         \color{black} \boldsymbol{X}^{s\dag}_{\hat{\bk},\hat{\bq}}
         \color{olive} \c2{\boldsymbol\Psi}_\bk(\tau) \,
         \color{olive} \c2{\boldsymbol\Psi}^\dag_{\bk}(\tau') \,
         \color{black} \boldsymbol{X}^s_{\hat{\bk},\hat{\bq}}
         \color{red}   \c1{\boldsymbol\Psi}_{\bk-\bq}(\tau')
    } \Big\rangle \Big].
\end{align}
In analogy to the calculations in \cref{Sec:rho-1} we find
\begin{align}
    \varrho_2(\tau) &= -\frac{1}{\pi} \frac{1}{a^4(\tau)} \int \! k^4 dk \! \int\!q^2 dq \!
                       \int \! d\theta \, \sin^3\theta \, \text{Im}\Big[
                           \int_{\tau^{+}_i}^\tau d\tau' \,
                           \big\langle \mathrm{h}_{q}(\tau) \, \mathrm{h}^*_q(\tau') \big\rangle \,
                           e^{i(\omega+k)(\tau'-\tau)}\Big] \nonumber\\
     &  \times  (k+\omega) \Big[ 2 - \frac{\sin^2\theta((\omega+k)^2+q^2)}{2\omega(\omega+k-q\cos\theta)}\Big].
       \label{Eq:varrho-1-2}
\end{align}
Considering once again the limit $\tau \gg \tau_*$ where $ \mathrm{h}_{q}(\tau)  \propto a^{-1}(\tau)$, we find that the time integral is asymptotically damped like
\begin{align}
    \lim_{\tau\gg \tau_*} \int_{\tau^{+}_i}^\tau d\tau' \,
        \big\langle \mathrm{h}_{q}(\tau) \mathrm{h}^*_q(\tau') \big\rangle \,
        e^{i(\omega+k)(\tau'-\tau)}
    \propto \frac{1}{a(\tau)}.
\end{align}
Therefore we have
\begin{align}
    \frac{\varrho_2(\tau)}{\varrho_1(\tau)} \propto \frac{1}{a(\tau)}.
    \label{eq:rho2-rho1}
\end{align}
As a result, $\varrho_2(\tau)$ is damped faster than radiation and is therefore negligible in cosmolgy.

\subsection{Contribution of the Quartic Vertex}
\label{sec:quartic-vertex}

The energy density also receives contributions from the quartic interaction, \cref{Eq:L2-int}. We write these contributions as
\begin{align}
    \ev{\rho(\tau,\bx)}\!_{\mathcal{L}^{_{(2)}}_\text{int}} = \varrho_3 + \varrho_4,
\end{align}
where 
\begin{align}
	\varrho_3	= \ev{ \rho_{\psi}^{(0)}	\mathord{
		\begin{tikzpicture}[radius=1.5cm, scale=0.6, baseline=-0.65ex, very thick]
 		\draw[decoration={markings, mark=at position 0.3 with {\arrow[scale=-1.5, rotate=6]{stealth}}, mark=at position 0.6 with {\arrow[scale=-1.5, rotate=6]{stealth}}}, postaction=decorate] (0, 0) circle [];
        \draw (-1.5,0) node {\huge $\times$}; 
        ;
  			\begin{feynman}
				\vertex (A) at (1.125,0.975);
				\vertex (C) at (0.1,0);
				\coordinate (D) at (-1.5, 0);
				\vertex (B) at (1.5, 0); 
				\diagram*{
					(B) -- [gluon, red, bend right] (C)
     -- [gluon, red, bend right] (B),
			}; \filldraw[white] (0,0) circle [radius=0.22cm];
          \draw[red, very thick] (0,0) node {\large{$\boldsymbol{\otimes}$}};
   			\end{feynman} 
	\end{tikzpicture} } } ,
    \qquad \qquad 
    \varrho_4	 = \ev{ \rho^{(2)}_{\psi}\mathord{
		\begin{tikzpicture}[radius=1.5cm, scale=0.6, baseline=-0.65ex, very thick]
 		\draw[decoration={markings, mark=at position 0.3 with {\arrow[scale=-1.5, rotate=6]{stealth}}, mark=at position 0.6 with {\arrow[scale=-1.5, rotate=6]{stealth}}, mark=at position 0.98 with {\arrow[scale=-1.5, rotate=7]{stealth}}}, postaction=decorate] (0, 0) circle [];
        \draw (-1.5,0) node {\huge $\times$}; 
        ;
  			\begin{feynman}
				\vertex (A) at (1.125,0.975);
				\vertex (C) at (-0.1,0);
				\coordinate (D) at (-1.5, 0);
				\vertex (B) at (1.125, -0.975); 
				\diagram*{
					(C) --  [red, gluon, bend right] (D)  --  [red, gluon, bend right] (C)
			};
    \filldraw[white] (0,0) circle [radius=0.22cm];
          \draw[red, very thick] (0,0) node {\large{$\boldsymbol{\otimes}$}};
   			\end{feynman} 
	\end{tikzpicture} }}.
    \label{Eq:loop2}
\end{align}
Both diagrams include the bubble diagram involving the quartic vertex $V_{hh\psi\psi}$,
\begin{align}
    \mathcal{A}(\bk,\tau) \equiv \hspace*{-2ex} \mathord{ 
         \begin{tikzpicture}[baseline=-0.11cm]
            \begin{feynman}
                \diagram [layered layout, horizontal=b to c] {
                 a [particle={$k$}] -- [fermion ] b [dot,blue]
                -- [red, gluon, min distance=2.3cm  ] b 
                -- [fermion ] c,
                }; 
            \end{feynman}
        \end{tikzpicture} }
    \label{Eq:bubble-diag}
\end{align}
Expanding $h_{ij}$ in $\mathcal{L}^{(2)}_\text{int}$ in Fourier modes, we find
\begin{align}
   \mathcal{A}(\bk,\tau) & \propto  \sum_{s} \int \! dq^3 \, \bigg[
        \Big( -\gamma^0 \gamma^5 q_j \epsilon_{ln}^{~~j} \boldsymbol{e}^{s\dag}_{in}(\hat\bq) \,
                \boldsymbol{e}^s_{il}(\hat\bq)
              + i \gamma^i q_k \boldsymbol{e}^{s\dag}_{kj}(\hat\bq) \, \boldsymbol{e}^s_{ij}(\hat\bq) \Big)
        \ev{ \hat{h}^{\dag}_{s,\bq}(\tau)\hat{h}_{s,\bq}(\tau) }  \nonumber\\
     & -i \gamma^i \gamma^5 \epsilon^{~~j}_{in} \boldsymbol{e}^{s*}_{lj}(\hat\bq) \,    
            \boldsymbol{e}^{s}_{ln}(\hat\bq)
            \ev{ \hat{h}^{\dag}_{s,\bq}(\tau) \hat{h}'_{s,\bq}(\tau) } \bigg] .
    \label{Eq:Del2}
\end{align}
Using the explicit form of the GW polarization tensors, we find
\begin{align}
    q_j \epsilon_{ln}^{~~j} \boldsymbol{e}^{s\dag}_{in}(\hat{\bq}) \, \boldsymbol{e}^{s}_{il}(\hat{\bq})
        = -i sq
\an
    \boldsymbol{\sigma}_j \epsilon_{ln}^{~~j} \, \boldsymbol{e}^{s\dag}_{in}(\hat{\bq}) \, 
        \boldsymbol{e}^{s}_{il}(\hat{\bq}) = -i s \hat{\bq}.\vec{\boldsymbol{\sigma}}.
\end{align}
The two terms in the first line of \cref{Eq:Del2} are thus proportional to $\sum_s s \big\langle \hat{h}^{\dag}_{s,\bq}(\tau) \hat{h}_{s,\bq}(\tau) \big\rangle$, and the term in the second line is proportional to $\sum_s s \big\langle \hat{h}^{\dag}_{s,\bq}(\tau) \hat{h}'_{s,\bq}(\tau) \big\rangle$. We assume that the mechanism producing the stochastic GW background is parity-conserving, in which case the GWs are unpolarized, and these expectation values are zero. Consequently, the gravitational bubble diagram associated with the quartic vertex, \cref{Eq:bubble-diag}, vanishes. That implies $\varrho_3$ and $\varrho_4$ vanish, hence
\begin{align}
    \ev{\rho(\tau,\bx)}_{\mathcal{L}^{_{(2)}}_\text{int}} = 0.
\end{align}
Therefore, for unpolarized GWs, the quartic vertex $V_{hh\psi\psi}$ does not contribute to the energy density of fermions.

\subsection{Background Waves}
\label{Sec:BGW-method}

Before we proceed to further study \cref{Eq:cG1-C} -- which, as we have seen, is the only cosmologically relevant contribution to $\ev{\rho_\psi(\tau)}$ -- we need to address one subtlety that we have not discussed yet. This is the choice of the integration limits in $q$. Here, we argue that mainly GW modes with $q \gtrsim k$ are relevant. Intuitively, this means that fermions of momentum $k$ can only be created by gravitons of momentum $q \gtrsim k$.

\begin{figure}
    \centering
    \includegraphics[scale=0.45]{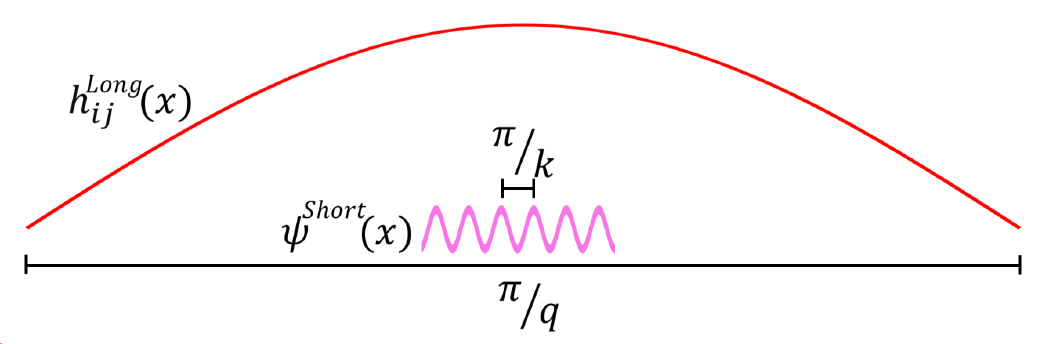}
    \caption{Long-wavelength GW modes with momentum $q$ act as a homogeneous field for a short-wavelength fermion mode with momentum $k$, where $q \ll k$.}
    \label{fig:Long-Short}
\end{figure}

The gravitational wave is a space-time fluctuation that appears as below in the perturbed metric
\begin{align}
    g_{ij}(t,\bx) \, dx^i dx^j = a^2 e^{2 h_{ij}(t,\bx)} dx^i dx^j.
\end{align}
The perturbation $h_{ij}(x)$ is gauge invariant (perturbatively) up to local diffeomorphisms. The main idea here is to separate the classical (sourced) GW spectrum into long-wavelength and short-wavelength contributions with respect to the wavelength of the fermion. For a fermion mode with momentum $k$, a GW perturbation with much smaller momentum, $q \ll k$ can be considered as spatially homogeneous background (see \cref{fig:Long-Short}). A long-wavelength GW mode $h^\text{long}_{ij}(t,\bx) \approx C_{ij}(t)$ where $ C_{ij}(t)$ is a homogenous tensor mode, can be recast as a change in coordinates \cite{Weinberg:2008zzc, Senatore:2012nq}:
\begin{align}
    h^{\text{long}}_{ij} \mapsto 0
    \quad \textmd{and} \quad
    x^i \mapsto x^{i'} = e^{h^{\text{long}}_{ij}} x^j.
\end{align}
In other words, it can be considered as a large gauge transformation. As a result, after this coordinate transformation, a short-wavelength fermion mode in the presence of long-wavelength GWs becomes a free fermion, i.e.,
\begin{align}
    \Psi_\text{short}(t,\bx) \big\rvert_{h_{ij}^{\text{long}}} = \Psi'(t,\bx') \big\rvert_0.
\end{align}
In terms of the fermion energy density, $\rho_{\psi}$, it can be written as
\begin{align}
    \langle \rho_{\psi,\text{short}}(x) \rangle \big\rvert_{h_{ij}^\text{long}}
        =  \langle \rho_{\psi,\text{short}}(x') \rangle \big\rvert_0 = 0.
\end{align}
In other words, long-wavelength GWs do not contribute to fermion production. Following this, and as an approximation, we only consider GWs with momentum $q > k$ when evaluating the momentum integrals. This approach is sometimes referred to as the background field method.

\subsection{Asymptotically Flat Spacetime and Astrophysical Sources of GWs}

While we have so far discussed the mechanism of GW-induced production of Weyl fermions only in the context of an expanding FLRW Universe, the careful reader may have realized that it should also work in asymptotically flat spacetimes. Thanks to the conformal symmetry of Weyl fermions at tree-level, we can go to the asymptotically flat case in our calculations by setting $a(\tau)=1$. The geometry is then given as
\begin{align}
    g_{\mu\nu} = \eta_{\mu\nu} + h_{\mu\nu}, \quad \qquad \lvert h_{\mu\nu} \rvert \ll 1,
\end{align}
where $\eta_{\mu\nu}$ is the Minkowski metric and $h_{\mu\nu}$ is a small perturbation which includes GWs. Crucially, though, these GWs are produced by \emph{local} astrophysical sources such as inspirals and mergers of compact objects. At distance $r$ from the source, the GW amplitude decreases as \cite{Maggiore:2007ulw}
\begin{align}
    h_s(t,r) \propto \frac{1}{r}
\end{align}
The physical reason underlying this behavior is the finiteness of the total GW energy and the asymptotic flatness of the spacetime \cite{Strominger:2017zoo}. Unfortunately, this makes fermion freeze-in induced by GWs from astrophysical sources negligible compared to the case of cosmological sources, which is the main focus of this work.

\subsection{Master Formula for GW-induced fermion production in an Expanding Universe}


To wrap up this section, let us recapitulate our finding. We have seen that four different gravitational one-loop diagrams contribute to the final fermion energy density, namely \cref{Eq:loop,Eq:loop2}, associated with the cubic vertex $V_{h\psi\psi}$ and the quartic vertex $V_{hh\psi\psi}$, respectively. We have computed the two diagrams from \cref{Eq:loop} separately in \cref{Eq:varrho-1-,Eq:varrho-1-2}, and have found that the second one gives a contribution that dilutes faster than radiation and is therefore unimportant in cosmology (see \cref{eq:rho2-rho1}). Moreover, in \Cref{sec:quartic-vertex}, we have shown that the diagrams from \cref{Eq:loop2} vanish for circularly unpolarized GW backgrounds. As a result, only the first diagram in \cref{Eq:loop} contributes to the final fermion energy density at gravitational 1-loop, therefore we arrive at the master formula
\begin{align}
    \langle \rho_{\psi}(\tau) \rangle &= \frac{1}{4\pi} \frac{1}{a^4(\tau)}
        \int_{\tau^+_i}^{\tau} d\tau' \int_{\tau^-_i}^{\tau} d\tau''
        \int \! q^2 dq \, \big\langle \mathrm{h}_{q}(\tau'') \mathrm{h}^*_{q}(\tau') \big\rangle
        \int \! k^4 dk \int \! d\theta \, \sin^3\theta \, e^{i(k+\omega)(\tau'-\tau'')}
                                                                \nonumber\\
    &\times (k+\omega) \Big[ 2 - \frac{\sin^2\theta((\omega+k)^2+q^2)}{2\omega(\omega+k-q\cos\theta)}\Big]  + c.c. \,.
    \label{Eq:rho-tot}
\end{align} 
One can also work out the pressure density of the produced fermions which is 
\begin{align}
    \langle P_{\psi}(\tau) \rangle = \frac13 \langle \rho_{\psi}(\tau) \rangle \propto \frac{1}{a^4(\tau)}.
    \label{eq:P}
\end{align}
This implies that as long as the fermions are effectively massless, they behave like radiation. Unlike the vacuum energy of fermions which is negative, the fermions produced via gravitational wave-induced freeze-in carry energy, i.e., $\langle \rho_{\psi}(\tau) \rangle > 0$.

In the following sections, we will further evaluate \cref{Eq:rho-tot}, using the background-wave approximation from \cref{Sec:BGW-method} as well as specific parameterizations of the GW background.

\section{Essentials of Stochastic Gravitational-Wave Backgrounds}
\label{Sec:GW}

Before proceeding further with the evaluation of the fermion abundance resulting from GW-induced freeze-in, we collect important formulas and conventions related to stochastic GW backgrounds, and we review the phenomenological parameterization we use to describe such backgrounds.

Stochastic GW backgrounds in the early Universe are commonly assumed to be statistically homogeneous and isotropic, unpolarized, and Gaussian. These assumptions are adopted also in the current work. For a comprehensive review of cosmological GW backgrounds see \cite{Maggiore:2007ulw, Maggiore:2018sht, Caprini:2018mtu}.

\subsection{Energy Density of Stochastic Gravitational Waves}
\label{Sec:GW-energy-density}

The GW energy density is related to the metric perturbation in the transverse--traceless gauge by the relation~\cite{Maggiore:2007ulw}
\begin{align}
    \rho_\text{gw}(t, \bx) = T^{00}_\text{gw}
        = \frac{1}{32\pi G} \big\langle \dot{h}_{ij}(t,\bx) \, \dot{h}^{ij}(t,\bx) \big\rangle.
\end{align}
Here $T^{00}_\text{gw}$ denotes the $(00)$ component of the energy-momentum tensor, and a sum over the spatial indices $i$, $j$ is implied. In Fourier space, the spectral energy density is correspondingly
\begin{align}
    \frac{d\rho_\text{gw}(\tau,q)}{d\ln q}
        &= 2 \pi q^3 \mpl^2
           \sum_s \langle \dot{\mathrm{h}}_{s,\bq}^*(\tau) \dot{\mathrm{h}}_{s,\bq}(\tau) \rangle
                                                  \nonumber\\
        &\simeq \frac{2 \pi q^5 \mpl^2}{a^2}
                \sum_s \langle \mathrm{h}_{s,\bq}^*(\tau) \mathrm{h}_{s,\bq}(\tau)\rangle,
    \label{Eq:Power-App}    
\end{align}
where $q$ is the comoving momentum (see \cref{Eq:GW}). In the second line, we have used the approximation valid for modes inside the horizon, $\dot{h}_{s,\bq} = h'_{s,\bq}/a \approx i q h_{s,\bq}/a$ (see ref.~\cite{Caprini:2018mtu} and references therein). In units of the critical density, $\rho_c = 3 H^2 / (8 \pi G)$, the spectral energy density is
\begin{align}
    \Omega_\text{gw}(\tau, q)
        \equiv \frac{1}{\rho_c(\tau)} \frac{d\rho_\text{gw}(\tau,q)}{d\ln q}
        =    \frac{2\pi q^5}{3 \mathcal{H}^2(\tau)}
             \mathcal{P}_\text{gw}(\tau,q),
    \label{Eq:Omega-vs-P}
\end{align}
where $\mathcal{H} = a H$ is the conformal Hubble parameter (see \cref{eq:rad-era-relations}), and where we have defined the GW power spectrum as
\begin{align}
    \mathcal{P}_\text{gw}(\tau,q) & \equiv
        \sum_s\langle \mathrm{h}^*_{s,\bq}(\tau)\mathrm{h}_{s,\bq}(\tau) \rangle \nonumber\\
        & = 2 \, \langle \mathrm{h}^*_q(\tau)\mathrm{h}_q(\tau) \rangle,
    \label{Eq:P-gw}
\end{align}
where in the second line we assumed GWs are circularly unpolarized (see \cref{Eq:power-GW}). Below, we will often need the GW power spectrum and dimensionless energy density today. We will refer to them as $\mathcal{P}_{\text{gw},0}(q)$ and $\Omega_{\text{gw}.0}(q)$, respectively. Big Bang Nucleosynthesis (BBN) imposes an upper bound on $\Omega_{\text{gw},0}$, namely \cite{Maggiore:2018sht}
\begin{align}
    \int_{f_\text{CMB}}^{\infty} \! d\ln q \; \Omega_{\text{gw},0}(q)
        \lesssim \num{3.5e-6}.
    \label{Eq:bbn}   
\end{align}
Here, $\Omega_{\text{gw},0}$ denotes the spectral energy density fraction today, and $f_\text{CMB} \simeq \SI{3e-17}{Hz}$ is the frequency associated with the Hubble radius at the CMB scale.

\subsection{The Gravitational Wave Spectrum}
\label{Sec:GW-spectrum}

The properties of cosmological GW backgrounds today depend on the production mechanism (dynamics) as well as the cosmic expansion history (kinematics). For phenomenological purposes, it is useful to separate the kinematics (which is simply the redshift from the production era to the present time) from the dynamics. We do this by writing
\begin{align}
    {\rm{h}}_{s,\bq}(\tau) = a^{-1}(\tau) \, \mathcal{T}(\tau,q) \, {\rm{h}}_{s\bq,0},
    \label{Eq:hsq-parameterization}
\end{align}
where ${\rm{h}}_{s\bq,0}$ is the gravitational wave amplitude today, the factor $a^{-1}(\tau)$ describes the redshift between conformal time $\tau$ and today, and $\mathcal{T}(\tau,q)$ is a transfer function which captures the dynamics of how the GW spectrum is built up over time. We denote the conformal times associated with the start and end of the GW production as $\tau_\text{in}$ and $\tau_*$  (settling-time) respectively. The transfer function should satisfy the boundary conditions
\begin{align}
    \mathcal{T}(\tau_\text{in},q) &= 0 , \\
    \mathcal{T}(\tau,q)           &\simeq e^{-iq\tau} \quad \text{for} \quad \tau > \tau_*,
\end{align}
such that for $\tau > \tau_*$, $\mathcal{T}(\tau,q)$ asymptotically approaches the free wave-like behavior. 

The spectral energy density of GWs before the settling-time $\tau_*$ is related to $\Omega_{\text{gw}}(\tau_*,q) $ as
\begin{align}
    \Omega_\text{gw}(\tau,q) \simeq \Big(\frac{a_*}{a(\tau)}\Big)^4
        \lvert \mathcal{T}(q,\tau) \rvert^2 \, \Omega_{\text{gw}}(\tau_*,q)
        \quad \text{for} \quad \tau < \tau_*.
    \label{Eq:Omega0-App0}
\end{align}
The relation to the GW energy density today, $\Omega_{\text{gw},0}$, in turn is
\begin{align}
    \Omega_{\text{gw},0}(q) = \Omega_{\text{rad},0} \,
                                  \frac{\rho_{\text{rad},*}}{\rho_{\text{rad},0}}
                                  \Big( \frac{a_*}{a_0} \Big)^4 \, \Omega_{\text{gw},*}(q),
\end{align}
with the obvious notation for the dimensionless ($\Omega_{\text{rad},0}$) and dimensionful ($\rho_{\text{rad},0}$) radiation energy density today and at the settling time ($\rho_{\text{rad},*}$). During the radiation-dominated era, we have $\rho_{\text{rad}} = (\pi^2 / 30) \, g(T) \, T^4$ where $g(T)$ is the effective number of relativistic degrees of freedom at temperature $T$. We can write $\mathcal{H}$ as
\begin{align}
    \label{Eq:cH-App}
    \mathcal{H} =\frac{1}{\tau}= \frac{\pi}{3} \sqrt{\frac{g(T)}{10}} \frac{T^2}{\mpl} \frac{1}{z},
\end{align}
with $z$ being the redshift. Numerically, we find 
\begin{align}
    \Omega_{\text{gw},0}(q) = \num{3.3e-5} 
                              \bigg( \frac{g(T_*)}{106.75} \bigg)
                              \bigg( \frac{g_{s}(T_*)}{106.75} \bigg)^{-\frac43} \Omega_{\text{gw},*}(q),
    \label{Eq:Omega0-App}
\end{align}
where we have used $\Omega_{\text{rad},0} \approx \num{4.2e-5} / h^2 \approx \num{8.5e-5}$, and $g_s(T)$ is the effective number of relativistic degrees of freedom appearing in the calculation of the entropy density. ($g_s(T)$ equals $g(T)$ as long as all relativistic species share the same temperature.) In this work, we are interested in $T_*$ above the electroweak scale, and we assume the dark fermions are never in thermal equilibrium with the hot plasma. This fixes $g(T_*) = g_s(T_*) = 106.75$. Now, using \cref{Eq:Omega0-App0} in \cref{Eq:Omega0-App}, we can write $\Omega_{\text{gw}}(\tau,q)$ at $\tau \leq \tau_*$ in terms of $\Omega_{\text{gw},0}(q)$ as
\begin{align}
    \Omega_\text{gw}(\tau,q) \simeq \num{3.0e4} 
        \Big( \frac{a_*}{a(\tau)} \Big)^4 \,
        \lvert \mathcal{T}(q,\tau) \rvert^2 \,
        \Omega_{\text{gw},0}(q) h^2
        \qquad \text{for} \quad \tau < \tau_*.
    \label{Eq:Omega0-App-2}
\end{align}
Using \cref{Eq:Omega0-App-2} in \cref{Eq:Omega-vs-P} we can also express the GW power spectrum numerically as
\begin{align}
    \mathcal{P}(\tau,q) \simeq  
 \frac{\num{1.8e5}}{4\pi} ~ \mathcal{H}^2(\tau) \, \mathcal{T}^2(q,\tau) \,
                               \frac{\Omega_{\text{gw},0}(q) h^2}{q^5},
    \label{Eq:Power-Omega-App}
\end{align}

\begin{figure}
    \centering
    \includegraphics[scale=0.5]{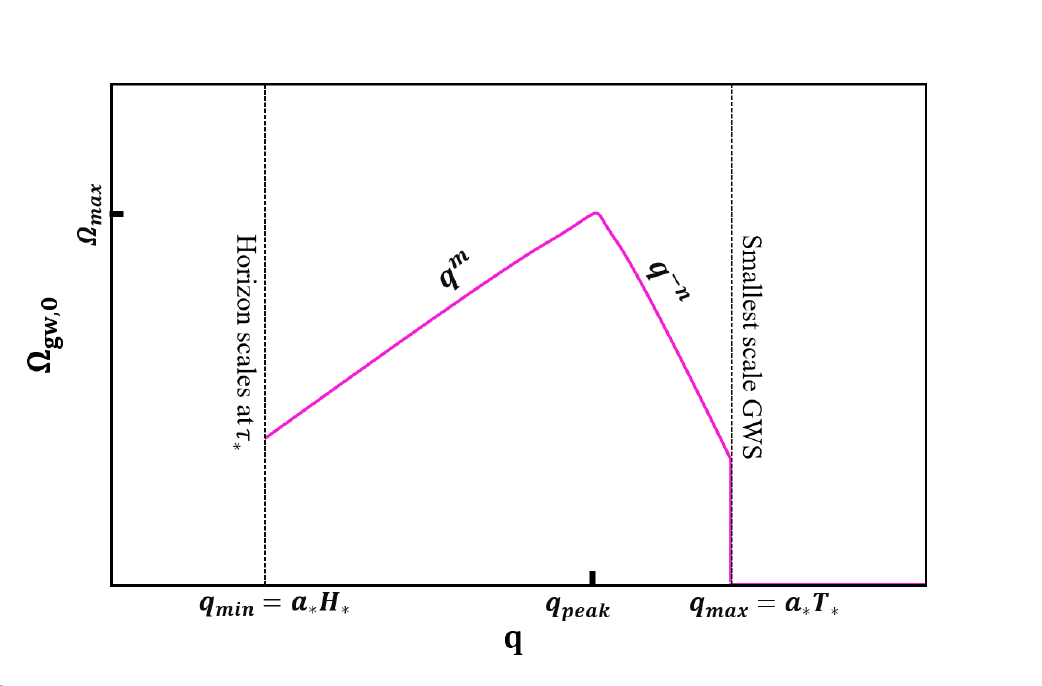}
    \caption{Schematic illustration of the spectral energy density of stochastic GWs generated by phase transitions and primordial magnetic fields. The spectrum has three distinct physical scales, $q_{\text{min}} \equiv q_{H}=a_*H_*$ (the horizon size at the time of phase transition), $q_{\text{max}}=a_*T_*$ (the smallest wavelength of the GW generated by the phase transition), and $q_{\text{peak}}$ (the wavenumber at the peak of the spectrum). Note this illustration is a simplified toy model, intended for conceptual clarity rather than exhaustive detail.}
    \label{Fig:Omega}
\end{figure}
 
We now need to model the transfer function $\mathcal{T}(\tau,q)$ and the spectral energy density today, $\Omega_{\text{gw},0}(q)$. For $\mathcal{T}(\tau,q)$, we use the parameterization
\begin{align}
    \mathcal{T}_{s}(q,\tau) \approx (1 - e^{-\pi \beta (\tau -\tau_\text{in}) }) \, e^{-iq\tau},
    \label{Eq:transfer-func}
\end{align}
where $\beta^{-1}$ is the time scale characterizing the duration of the cosmological process that sources GWs. The motivation for choosing this particular form is purely phenomenological. It is a simple way to model a signal that starts at $\tau = 0$ and levels out at $\tau > \beta^{-1}$ to take the form of a propagating wave. It is assumed that $\beta^{-1}$ is faster than the Hubble time, i.e.\ $\beta/\mathcal{H}_*>1$. In this work, the notation $\beta$ denotes the exponent combined with conformal time, similar to the convention used in \cite{Caprini:2018mtu}. It is worth noting that $\beta$ is sometimes defined differently, namely as a rate with respect to cosmic time instead of conformal time \cite{Maggiore:2018sht}. With this alternative definition, the transfer function would be proportional to $(1 - e^{-\pi \beta (t - t_\text{in}) })$ instead of $(1 - e^{-\pi \beta (\tau -\tau_\text{in})})$. Consequently, in the former, a dimensionless quantity is introduced as $\beta/\mathcal{H}_*$, while in the latter it is $\beta/H_*$.

The spectral energy density of the stochastic GW background depends on the production mechanism and typically needs to be determined using sophisticated simulations.  In many cases, however, the results can be well described by simple analytic fitting functions.
\begin{enumerate}[i)]
    \item {\bf Phase transitions} generate a GW spectrum that can be fitted by a broken power law \cite{Caprini:2009yp, Durrer:2010xc} (see \cref{Fig:Omega}):
    \begin{align}
        \Omega_{\text{gw},0}(q) \approx
            \begin{cases}
                \Omega_\text{peak} \, \big( \frac{q}{q_\text{peak}} \big)^m
                    & \text{for} \quad q_\text{min} < q < q_\text{peak} , \\
                \Omega_\text{peak} \, \big( \frac{q}{q_\text{peak}} \big)^{-n}
                    & \text{for} \quad q_\text{peak} < q < q_\text{max} ,
                    \end{cases}
        \label{Eq:Omega-PT}
    \end{align}
    where $\Omega_\text{peak}$ is the value of the spectrum at the peak and $q_\text{peak}$ is the momentum at the peak. This spectrum has three physical scales. The low-frequency cutoff $q_\text{min} = a_* H_*$ is the horizon size, while the high-frequency cutoff $q_\text{max} \approx a_* T_*$ is a characteristic (physical) scale of the system associated with the smallest scales beyond which the source decays.  The spectral indices are $m \approx 3$ and $n \sim (1-4)$; the exact value of $n$ varies across different scenarios with different degrees of coherence in time \cite{Caprini:2009yp, Durrer:2010xc}.\footnote{If the anisotropic stress spectrum of the source decays as $q^{-\gamma}$, then, for perfect temporal coherence, the gravitational wave energy density decays as $q^{-(1+\gamma)}$ at high frequencies. If the source is partially coherent, i.e., for about one wavelength, the GW energy density decays as $q^{-(\gamma-1)}$. Finally, a totally incoherent source leads to a much softer GW energy spectrum, falling as $q^{-(\gamma-3)}$ \cite{Caprini:2009yp, Durrer:2010xc}.} The BBN bound from \cref{Eq:bbn} imposes the condition 
    \begin{align}
        \Omega_\text{peak} \lesssim \num{3.5e-6} \frac{m \, n}{m+n}.
    \end{align} 

    \item {\bf Primordial magnetic fields} can also source a stochastic gravitational wave background \cite{Caprini:2018mtu}. Similar to the case of phase transitions, the resulting GW energy spectrum can be modeled as a broken power law \cite{RoperPol:2022iel}. For the purpose of our analytical approximations, we will therefore use the parameterization from \cref{Eq:Omega-PT} also for GWs from primordial magnetic fields.
    
    \item {\bf Reheating and gauge reheating} produce GW spectra with one or multiple peaks, and with a fast drop-off towards high frequencies \cite{Adshead:2019igv, Figueroa:2022iho}. The detailed shape of the spectrum varies model by model. Since our momentum integrand rapidly decays for the decreasing portions of the spectrum, we can simplify the approximation by focusing solely on the initial peak, effectively represented by \cref{Eq:Omega-PT}.

    \item {\bf Cosmic strings} typically lead to a GW spectrum featuring a flat plateau which decays at very high frequencies \cite{Hindmarsh:1994re, Auclair:2019wcv}. This scenario is not well described by \cref{Eq:Omega-PT}, so we leave it for future work involving more advanced modeling.
\end{enumerate}

\subsection{Temporal Coherence of GWs}
\label{Sec:GW-coherence}

The unequal-time two-point correlation function of the GW background, which appears in \cref{Eq:rho-tot}, is considerably more challenging to determine than the GW power spectrum. In particular, it requires information about the temporal coherence of the source. We will desccribe the latter using simple phenomenological models \cite{Caprini:2009fx, Caprini:2009yp}, which fall into three distinct categories: (i) fully incoherent in time, (ii) fully coherent in time, and (iii) an intermediate regime characterized by a finite coherence time.  For instance, bubble collisions during a first-order phase transition can be modeled as fully coherent, i.e., deterministic in time. Turbulence and magnetic fields, on the other hand, are partially coherent \cite{Caprini:2009yp}.  The two-point correlation function can be written as \cite{Scully_Zubairy_1997}
\begin{align}
    \ev{\mathrm{h}_{q}^{*}(\tau'') \mathrm{h}_{q}(\tau')}
      = \gamma_q\big(\lvert \tau'-\tau''\rvert\big)
        \sqrt{\ev{\lvert \mathrm{h}_{q}(\tau')\rvert^2}\ev{\lvert \mathrm{h}_{q}(\tau'') \rvert^2} }
        ,
    \label{Eq:coherent}
\end{align}
where $\gamma_q\big(\lvert \tau'-\tau''\rvert\big)$ is a function of $\lvert \tau'-\tau''\rvert$ which specifies the degree of temporal coherence. For the sake of our analytical estimate, we specifically focus on cases (i) and (ii) above.
\begin{itemize}
    \item For fully incoherent GWs, the coherence function is 
    \begin{align}
        \gamma_q (\big\lvert \tau'-\tau'' \big\rvert) = \Delta\eta \, \delta(\tau'-\tau''),\label{Eq:tot-incoh-def}
    \end{align}
    where the characteristic coherence time $\Delta\eta$ is much shorter than the dynamical time scales in the system, i.e. $\Delta\eta \ll \beta ^{-1}$ \cite{Caprini:2009yp}.
    
    \item For fully coherent GWs, we have instead 
    \begin{align}
        \gamma_q (\big\lvert \tau'-\tau'' \big\rvert) = 1.
    \end{align}
\end{itemize}

\section{Analytical Estimates}\label{Sec:analytical}

Having worked out the master formula for the energy density of the fermions generated by gravitational wave-induced freeze-in in \cref{Sec:fermion-1-loop}, and having discussed our phenomenological models for the primordial GW background in \cref{Sec:GW}, we are now ready to find an analytical estimate for the fermion energy density. Notably, we wish to evaluate \cref{Eq:rho-tot} for $\ev{\rho_\psi(\tau)}$. A critical part of this expression is the time integral
\begin{align}
    {\rm{T}}(\tau; \bk,\bq) \equiv \int_{\tau^+_i}^{\tau} d\tau' \int_{\tau^-_i}^{\tau} d\tau'' \,
         {\rm{Re}} \Big[\big\langle \mathrm{h}_{q}(\tau'') \, \mathrm{h}^*_{q}(\tau') \big\rangle
                        e^{i(k+\omega)(\tau'-\tau'')} \Big],
    \label{Eq:t-function}
\end{align} 
which depends on the unequal-time correlator of the GWs, and therefore on their degree of temporal coherence. In the following, we will consider the two extreme cases introduced in \cref{Sec:GW-coherence}: fully coherent and fully incoherent GWs.

\subsection{Temporally incoherent GW background}
\label{Sec:incoh-GW}

According to \cref{Eq:coherent} and \cref{Eq:tot-incoh-def}, for a GW background that is totally incoherent in time, and after using the transfer function \cref{Eq:transfer-func}, we can write
\begin{align}
    \big\langle \mathrm{h}_q(\tau'') \, \mathrm{h}^*_q(\tau') \big\rangle
        & \approx \Delta\eta \, \delta(\tau' - \tau'') \ev{|\mathrm{h}_q(\tau')|^2} 
        \nonumber\\
        & \approx \frac12 \frac{\Delta\eta}{a^2(\tau')} (1-e^{-\pi\beta(\tau'-\tau_{\text{in}})})^2 \, \delta(\tau' - \tau'') \mathcal{P}_{\text{gw},0}(q).
\end{align}
The factor $1/a^2(\tau)$ arises when we express $\mathcal{P}_{\text{gw}}(\tau,q)$ in terms of $\mathcal{P}_{\text{gw},0}(q)$.
To evaluate the time integral from \cref{Eq:t-function}, ${\rm{T}}(\tau; \bk,\bq)$, we need the explicit $\tau$ dependence of $a(\tau)$ which, in the radiation era, can be written as
\begin{align}
    \frac{a(\tau)}{a_0} = \frac{\tau}{\tau_* z_*},
    \label{Eq:a-tau-relations}
\end{align}
with $z_*$ the redshift at $\tau_*$. Using moreover that $\mathcal{H} = a H = 1/\tau$ during radiation domination, we then find
\begin{align}
    \rm{T}(\tau; \bk,\bq) &= \frac{1}{2}
        \Delta\eta \, z_*^2 \int_{\tau_i}^{\tau} d\tau' \,
        \Big( \frac{\tau_*}{\tau'} \Big)^2
        \big( 1 - e^{-\pi\beta(\tau'-\tau_{\text{in}})} \big)^2 \,
        \mathcal{P}_{\text{gw},0}(q) \nonumber\\
    &\approx  \frac32 \ln2 \, z_*^2 \, \frac{\Delta\eta \, \beta}{\mathcal{H}_*^2} \,
        \frac{H_0^2}{q^5} \Omega_{\text{gw},0},
        \label{eq:T-incoh}
\end{align}
where in the second line we have considered the limits $\tau_{\text{in}} \ll \beta^{-1}$, $\beta \gg \mathcal{H}_*$, and  $\tau \gg \tau_*$, thereby effectively extending the integration region to the interval $(0, \infty)$. We have required that GW emission proceeds fast compared to Hubble expansion. We have also used \cref{Eq:Omega-vs-P} to express $\mathcal{P}_{\text{gw},0}(q)$ in terms of $\Omega_{\text{gw},0}$. Inserting the above in \cref{Eq:rho-tot}, we find
\begin{align}
    \langle \rho_{\psi}(\tau) \rangle = 
        \frac{3 \ln2}{8\pi} \frac{ z_*^2 \, \Delta\eta \, \beta}{a^4(\tau)}
        \frac{H_0^2}{\mathcal{H}_*^2}
    \int \! dq \, \frac{1}{q^3} \Omega_{\text{gw},0}(q)
    \int_{q_H}^q \! k^4 dk \, \mathcal{A}(k,q), 
    \label{Eq:rho-tot-incoh-1}
\end{align} 
in which $\mathcal{A}(k,q) $ is the angular integral
\begin{align}
    \mathcal{A}(k,q) &\equiv 2 \int \! d\theta \, \sin^3\theta \,
      (k+\omega) \bigg[
          2 - \frac12 \frac{(\omega+k)^2 + q^2}{\omega(\omega + k - q\cos\theta)} \, \sin^2\theta 
      \bigg]
                                       \nonumber\\
    &=\frac{8 q}{3}+\frac{16 k}{5}+\frac{16 k^2}{15 q}+\frac{8 k^4}{35 q^3}-\frac{32 k^6}{315 q^5}.
\end{align}
The leading factor of two arises from the complex-conjugate term in \cref{Eq:rho-tot}. Note that in \cref{Eq:rho-tot-incoh-1}, we have restricted the range of the integral over fermion momenta $k$ to the region $k < q$, following the arguments given in \cref{Sec:BGW-method}. The lower integration boundary is $q_H = a(\tau_*) H(\tau_*)$, corresponding to horizon-size GW modes at the settling time. Evaluating the integral over $k$ leads to
\begin{align}
    \int^q_{q_H} k^4dk \, \mathcal{A}(k,q)
        & = \frac{856 q^6}{693}-\frac{8 q_H^5 q}{15}-\frac{8 q_H^6}{15}-\frac{16 q_H^7}{105 q}-\frac{8 q_H^9}{315 q^3}+\frac{32 q_H^{11}}{3465 q^5}.
\end{align}
To be able to evaluate the integral over GW momenta $q$, we use the broken power law spectrum in \cref{Eq:Omega-PT} for $\Omega_{\text{gw},0}(q)$, and we split the integral into the region below $q_\text{peak}$ and the region above. Using moreover $q_\text{peak} \gg q_H$, we have
\begin{align}
    \int_{q_H}^{q_\text{peak}} \frac{dq}{q^3}
        \Big(\frac{q}{q_\text{peak}}\Big)^m
        \int^q_{q_\text{H}} k^4 dk \, \mathcal{A}(k,q)
    \approx \frac{856}{693} \frac{q_{\text{peak}}^4}{m+4},
\end{align}
and 
\begin{align}
    \int_{q_\text{peak}}^{q_\text{max}} \frac{dq}{q^3}
        \Big(\frac{q}{q_\text{peak}}\Big)^{-n}
        \int^q_{q_\text{H}} k^4dk \, \mathcal{A}(k,q)
    \approx \begin{cases}
                \frac{856}{693} \frac{q_\text{peak}^4}{4-n} \big[ \big(
                    \frac{q_\text{max}}{q_\text{peak}} \big)^{4-n} - 1 \big]
                                         & n \neq 4,\\
                \frac{856}{693} q_\text{peak}^4 \ln\Big(\frac{q_\text{max}}{q_\text{peak}}\Big) 
                                         & n = 4.
            \end{cases}
\end{align}
Putting everything together, the final energy density of Weyl fermions is then given as
\begin{align}
    \langle \rho_{\psi}(\tau) \rangle
        = \Big( \frac{q_\text{peak}}{a(\tau)} \Big)^4 \,
          \Big( \frac{H_0}{\mathcal{H}_*} \Big)^2 \, z_*^2 \,
          \mathcal{C}_\text{incoh} \, \Omega_\text{peak}, 
    \label{Eq:rho-tot-incoh-2}
\end{align} 
where for realistic scenarios in which $q_\text{H}\ll q_\text{peak} \ll q_\text{max}$, $\mathcal{C}_{\text{incoh}}$ is 
\begin{align}
    \mathcal{C}_\text{incoh} \approx \frac{107 \ln2}{231\pi} \,\Delta\eta \, \beta
        \begin{cases}
            \frac{1}{4-n} \big(\frac{q_\text{max}}{q_\text{peak}} \big)^{4-n}
                                          & n<4,\\
            \ln\Big( \frac{q_\text{max}}{q_\text{peak}} \Big)
                                          & n=4, \\
            \frac{(n+m)}{(4+m)(n-4)}      & n>4.
        \end{cases}
    \label{Eq:Cmn-incoh}
\end{align}
We observe that the behavior of $\rho_{\psi}(\tau) \propto \mathcal{C}_\text{incoh}$ is governed by the slope of the GW spectrum at high frequencies, $n$. $\mathcal{C}_\text{incoh}$ is an order-one number for $n \geq 4$, but becomes notably large for $n < 4$.

\subsection{Temporally Coherent GW Background}\label{Sec:coh-GW}

\begin{figure}
    \centering
    \includegraphics[width=0.5\textwidth]{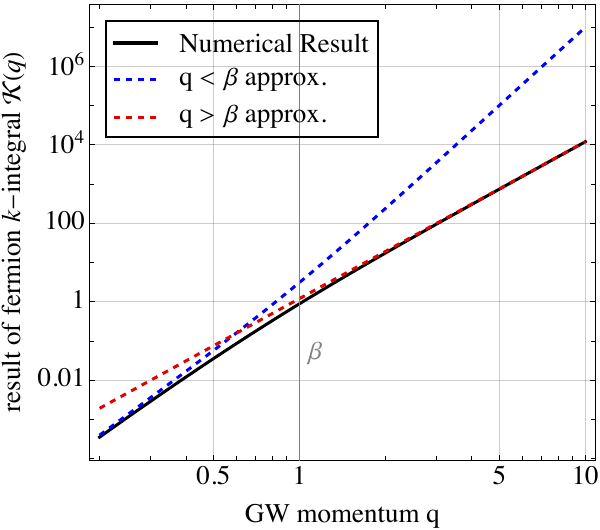}
    \caption{Comparison of the approximate expressions for the integral $\mathcal{K}(q)$ from \cref{Eq:K-q-2} to the numerical value of $\mathcal{K}(q)$. We have chosen $\beta = 1$ here.}
    \label{fig:K-q-approx}
\end{figure}

The computation of the integrals for a temporally coherent GW background is more involved. As discussed in \cref{Sec:GW-coherence}, the GW background in this case can be factorized as
\begin{align}
    \langle h^*_q(\tau'') h_q(\tau') \rangle
        =      \sqrt{ \ev{|h_q(\tau')|^2} \ev{|h_q(\tau'')|^2} },
\end{align}
Using this expression in \cref{Eq:t-function} and writing the GW power spectrum at conformal time $\tau$ in terms of the power spectrum today with the help of \cref{Eq:hsq-parameterization,Eq:transfer-func}, we find
\begin{align}
    \rm{T}(\tau; \bk,\bq) &= \frac12
        \frac{z_*^2}{\mathcal{H}_*^2} \, \bigg\vert
            \int_{\tau_i}^\tau \frac{d\tau'}{\tau'}
            \big( 1 - e^{-\pi\beta\tau'} \big) e^{i(\omega+k+q)\tau'}   \bigg\vert^2
            \, \mathcal{P}_{\text{gw},0}(q) \\
    &\approx \frac12
        \frac{z_*^2}{\mathcal{H}_*^2}
        \ln\Big[1 -\frac{i\pi\beta}{\omega+q+k}\Big]
        \ln\Big[1 +\frac{i\pi\beta}{\omega+q+k}\Big] \,
        \mathcal{P}_{\text{gw},0}(q).
    \label{eq:T-coh}
\end{align} 
The prefactor $z_* / (\mathcal{H}_* \tau')$ arises from the relation between $a$ and $\tau$, \cref{Eq:a-tau-relations}. In the second line, we have assumed $\beta \tau_* \gg 1$. The fermion energy density is then
\begin{align}
    \langle \rho_{\psi}(\tau) \rangle &= 
        \frac{2}{8\pi} \frac{1}{a^4(\tau)} \frac{z_*^2}{\mathcal{H}_*^2}
        \int \! q^2 dq \, \mathcal{P}_{\text{gw},0}(q)
        \int_{q_H}^q \! k^4 dk \, \int \! d\theta \, \sin\theta   (k + \omega) \, \mathcal{D}(\hat{\bq},\hat{\bk}) \nonumber\\
   & \times  \ln\Big[1 - \frac{i\pi\beta}{\omega+q+k}\Big]
        \ln\Big[1 + \frac{i\pi\beta}{\omega+q+k}\Big],
 \label{Eq:rho-tot-coh}
\end{align}
with $\mathcal{D}(\hat{\bq},\hat{\bk})$ from \cref{Eq:C-qk}. The leading factor of two is again due to the complex-conjugate term in \cref{Eq:rho-tot}. The next step is to carry out the integrals over $\theta$ and $k$,
\begin{align}
    \mathcal{K}(q) \equiv
      2  \int_{q_\text{H}}^q \! dk \, k^4 \int \! d\theta \sin\theta
     (k + \omega)  \, \mathcal{D}(\hat{\bq},\hat{\bk}) \,
        \ln\Big[1 - \frac{i\pi\beta}{\omega+q+k}\Big]
        \ln\Big[1 + \frac{i\pi\beta}{\omega+q+k}\Big],
    \label{Eq:K-q-1}
\end{align}
which can be approximated as
\begin{align}
    \mathcal{K}(q) \approx 
    \begin{cases} 
        \big( 3.1 - 0.077 \ln\frac{\beta }{q} + 1.2 \ln^2\frac{\beta }{q} \big) q^6
                 &\text{for} \quad \mH_* \lesssim q \lesssim \beta,\\
        \frac{38\pi^2}{315} \beta^2 q^4
                 &\text{for} \qquad \quad ~~ q\gtrsim \beta.
    \end{cases} 
    \label{Eq:K-q-2}
\end{align}
In the first case (low $q$), we have approximated $\ln\big[1 \pm \frac{i\pi\beta}{\omega+q+k}\big] \simeq \ln\big[\pm \frac{i\pi\beta}{\omega+q+k}\big]$, while in the second case (high $q$), we haved used $\ln\big[1 \pm \frac{i\pi\beta}{\omega+q+k}\big] \simeq \pm \frac{i\pi\beta}{\omega+q+k}$. Note that the exact coefficients in the low-$q$ case are rather lengthy, therefore we here give only their approximate numerical values to two significant digits. The goodness of these approximations is illustrated in \cref{fig:K-q-approx}, where we compare them to the result of a fully numerical evaluation of $\mathcal{K}(q)$.

Using the broken power law GW spectrum from \cref{Eq:Omega-PT}, the relation \cref{Eq:Omega-vs-P} between $\mathcal{P}_{\text{gw},0}(q)$ and $\Omega_{\text{gw},0}(q)$, as well as $q_\text{peak} \approx \beta$, we can now evaluate the integral over $q$. For the rising part of the GW spectrum (below the peak frequency), we find
\begin{multline}
    \int_{q_\text{H}}^{q_\text{peak}} q^2 dq \, \mathcal{P}_{\text{gw},0}(q) \, \mathcal{K}(q) \\
    \approx \frac{3 H_0^2 }{2\pi} \frac{q^4_\text{peak} \Omega_\text{peak}}{m+4} \,
    \bigg[ 3.1 
         + \Big( \frac{2.5}{4+m} - 0.077 \Big) \ln\frac{\beta}{q_\text{peak}}
         + 1.2 \ln^2\frac{\beta}{q_\text{peak}}
    \bigg],
\end{multline}
where we have neglected a very weak dependence on $m$ in the first term in square brackets, and we have again evaluated the somewhat lengthy numerical prefactors to two significant digits. The falling part of the GW spectrum (above the peak) gives
\begin{multline}
    \int^{q_\text{max}}_{q_\text{peak}} q^2 dq \, \mathcal{P}_{\text{gw},0}(q)
        \, \mathcal{K}(q)
    \approx \frac{3H_0^2 }{2\pi} \frac{38\pi^2}{315} q_\text{peak}^4 \Omega_\text{peak}
            \Big( \frac{\beta}{q_\text{peak}} \Big)^2 \\
            \times
            \begin{cases} 
                \frac{1}{n-2} \big[
                    1 - (q_\text{max}/q_\text{peak})^{2-n} \big]
                                         & \text{for} \quad n \neq 2,\\
                \ln\frac{q_{\text{max}}}{q_\text{peak}}
                                         & \text{for} \quad n = 2.
            \end{cases}
\end{multline}
This finally leads to the following expression for the energy density of Weyl fermions:
\begin{align}\label{Eq:rho-coh-}
    \langle \rho_{\psi}(\tau) \rangle \approx
         \, \frac{q^4_\text{peak}}{a^4(\tau)}
        \frac{H_0^2}{\mathcal{H}_*^2} \, z_*^2 \, \mathcal{C}_\text{coh} \Omega_\text{peak} ,          
\end{align}
in which $\mathcal{C}_\text{coh}$ is given as
\begin{multline}
    \mathcal{C}_{\text{coh}} \approx
        \frac{1}{m+4} \, \bigg[ 0.059 
            + \Big( \frac{0.047}{m+4} - 0.0015 \Big) \ln\frac{\beta}{q_\text{peak}}
            + 0.023 \ln^2\frac{\beta}{q_\text{peak}}
        \bigg]
        \\
      + 0.023 \Big( \frac{\beta}{q_\text{peak}} \Big)^2
            \times
            \begin{cases} 
                \frac{1}{n-2} \big[ 1 - (q_\text{peak}/q_\text{max})^{n-2} \big]
                                         & \text{for} \quad n \neq 2,\\
                \ln\frac{q_{\text{max}}}{q_\text{peak}}
                                         & \text{for} \quad n = 2.
            \end{cases}    
    \label{Eq:Cmn-coh}
\end{multline}

\section{Dark Matter Relic Density}
\label{Sec:relic}

\begin{figure}[t]
    \centering
    \includegraphics[scale=0.5]{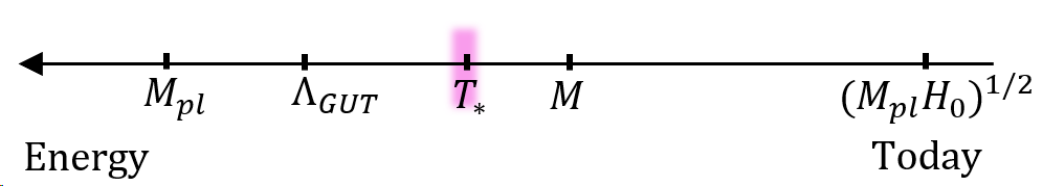}
    \caption{The hierarchy of scales in GW-induced freeze-in of Weyl fermions. $T_*$ is the temperature at the end of the cosmic event instigating the production of gravitational waves, and $M$ is the mass of the dark (Majorana/Dirac) fermions. To form the dark matter in the Universe, the fermions are assumed to be effectively massless during the epoch where they are produced, but carry a non-zero mass today.}
    \label{fig:hierarchy}
\end{figure}

Having analytically estimated the energy density of Weyl fermions generated by a stochastic GW background, we now explore the possibility that these fermions form the dark matter in the Universe. Clearly, this requires that they have a non-zero mass $M$ today, in spite of being effectively massless at the time of production, see \cref{fig:hierarchy}. Their mass today could either come from a tree-level mass term that is negligible at $T_*$ (i.e. $M < H_*$), but not today, or they could be exactly massless early on and acquire a mass later through a Higgs mechanism or through a confining phase transition.

We will consider only fermion production, neglecting annihilation. This is justified even in our minimal, CP-symmetric, setup because our fermions are feebly interacting and never come into thermal equilibrium with the SM or with each other. Therefore, fermion--antifermion annihilation is unimportant.

\subsection{GW-Induced Freeze-In}

If we call $\tau_M$ the conformal time at which $T \sim M$, then at $\tau > \tau_M$, the fermion energy density is (see \cref{Eq:rho-tot-incoh-2} and \eqref{Eq:rho-coh-})%
\footnote{Notice that \cref{Eq:rho-tot-incoh-2} and \eqref{Eq:rho-coh-} exhibit the expected $\mpl^{-2}$ suppression through $\Omega_\text{peak}\propto \frac{1}{\mpl^2}$. As a result, the backreaction of relativistic fermion production onto the GW background is negligible, i.e. $\rho_{\psi} / \rho_\text{gw} \sim \mathcal{C} q_\text{peak}^4 / (\mathcal{H}_*^2 \mpl^2) \ll 1$.}
\begin{align}
    \langle \rho_\psi(\tau) \rangle \big|_{\tau > \tau_M}
        = \frac{a_0}{a(\tau_M)} 
          \frac{q^4_\text{peak}}{a^3(\tau)} \frac{H_0^2}{\mathcal{H}_*^2} \,
          z_*^2 \, \mathcal{C} \, \Omega_\text{peak},
          \label{eq:rho-M}
\end{align}
where $\mathcal{C}$ is given in \cref{Eq:Cmn-incoh} and \cref{Eq:Cmn-coh} for incoherent and coherent GW backgrounds, respectively.  Using the conservation of the entropy density $s(\tau) = (2\pi^2 / 45) g_*(\tau) T^3(\tau)$ in comoving momentum, $s \, a^3 = \text{const}$, we can substitute
\begin{align}
    \frac{a_0}{a(\tau_M)} = \bigg( \frac{M}{T_0}\bigg)
                            \bigg( \frac{g_{*}(\tau)}{g_{*,0}}\bigg)^{\frac13} ,
\end{align}
where $T_0 = \SI{2.3e-4}{eV}$ is the temperature of the universe today, $g_{*,0}= 3.38$ is the effective number of relativistic degrees of freedom today, and $g_*(\tau)$ is the corresponding number at $\tau = \tau_M$. We use $g_*(\tau) = 106.75$, the value above the electroweak scale in the Standard Model. Using moreover \cref{Eq:cH-App} for $\mathcal{H}_*$ and dividing by the critical density $\rho_c = 3 \mpl^2 H_0^2$, we find the total fractional dark fermion energy density today
\begin{align}\label{Eq:OmegaD--}
    \Omega_{\psi,0}
        &= \frac{ \pi^2}{270} \, g_*(\tau_*)  \, \mathcal{C} \,
           \bigg( \frac{M}{T_0} \bigg)
           \bigg( \frac{g_*(\tau_*)}{g_{*,0}}\bigg)^{\frac13}
           \Big(\frac{q_\text{peak}}{\mathcal{H}_*}\Big)^4
           \Big(\frac{T_*}{\mpl}\Big)^4 \, \Omega_\text{peak}     \\
        &\simeq 0.36 \times \mathcal{C}
           \bigg( \frac{M}{T_*} \bigg)
           \bigg( \frac{q_\text{peak} / \mathcal{H}_*}{100} \bigg)^4
           \bigg( \frac{g_*(\tau_*)}{106.75} \bigg)^{4/3}
           \bigg( \frac{T_*}{\SI{3 e11}{GeV}} \bigg)^5
           \bigg( \frac{\Omega_\text{peak}}{\num{e-6}} \bigg).
    \label{Eq:OmegaD}
\end{align}
Note that GW-induced freeze-in is a non-thermal process and the appearance of $T_*$ is due to the redshift and Hubble parameter of the phase transition in \cref{eq:rho-M}. Now, let us pause for a qualitative discussion on the scaling and behavior of the final relic density. It can be more clearly understood by referring to the master formula in \cref{Eq:rho-tot}. We have 2 time integrals each giving a factor of $\frac{1}{\mathcal{H}_*}$ (duration of time integration). The GWs 2-point function can be written as $P_\text{gw}\propto \frac{\rho_\text{gw}}{ \mpl^2}$. There are also 4 powers of $q_\text{peak}$,  arising from the two momentum integrals. Substituting these into \cref{Eq:rho-tot}, we find $\rho_{\psi}/\rho_c \propto \frac{q_\text{peak}^4}{\mpl^4}\frac{1}{\mathcal{H}_*^2}(\frac{\rho_\text{gw}}{H_0^2})$. 
Next, when counting $\mathcal{H}_*$ as $T_*^2/\mpl$, it is essential to also count $q_\text{peak}$, since our scaling is $q_\text{peak}/\mathcal{H}_* \sim 10^{2}-10^{3}$. 
Additionally, $\rho_\text{gw}$ is the energy density of stochastic classical gravitational waves, specified by the early universe processes that generate them. After the fermions become non-relativistic, a factor of $\frac{M}{T_0}$ arises. Combining these factors, the power counting of $\mpl$
in $\Omega_{\psi,0}$ is expressed as $\Omega_{\psi,0} \sim \frac{M}{T_0}\frac{T_*^4}{\mpl^6} \frac{\rho_\text{gw}}{H_0^2}$, as shown in \cref{Eq:OmegaD--}.

\begin{figure}
    \centering
    \begin{tabular}{cc}
        \includegraphics[width=0.5\textwidth]{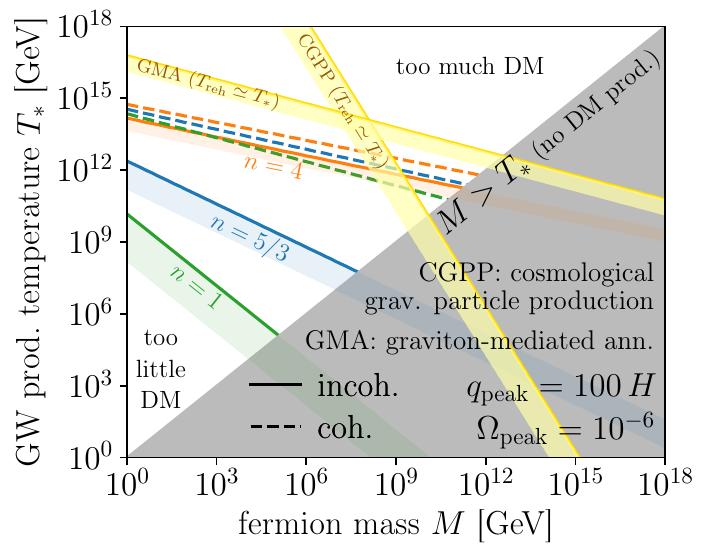} &
        \includegraphics[width=0.5\textwidth]{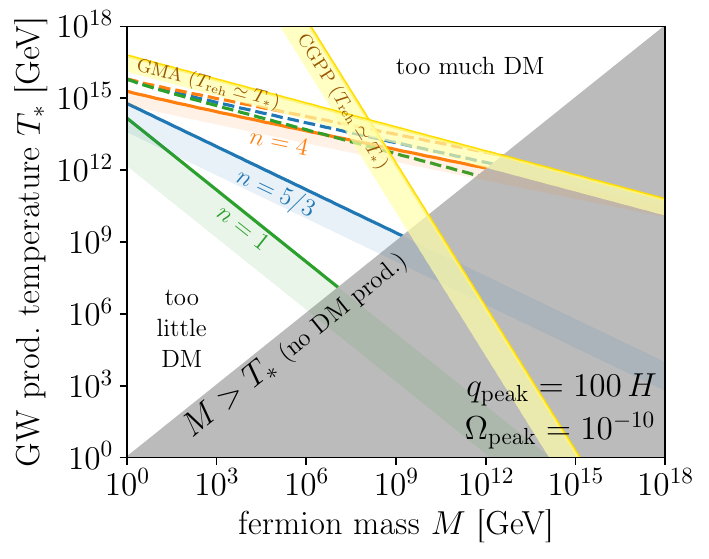} \\
        \includegraphics[width=0.5\textwidth]{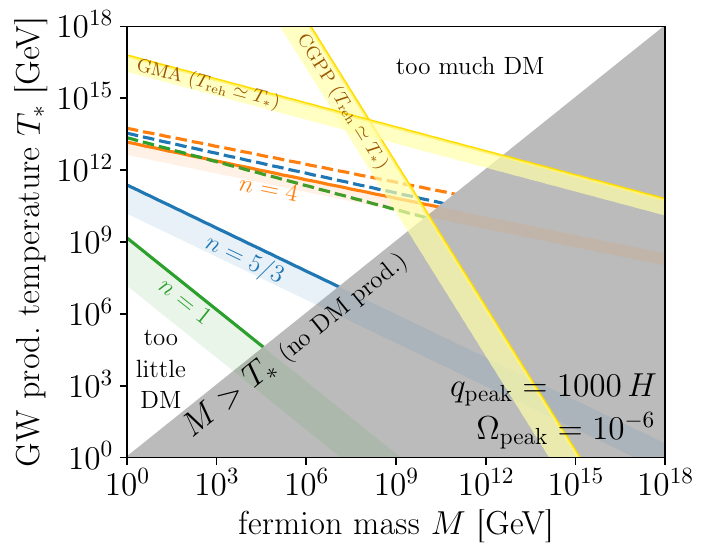} &
        \includegraphics[width=0.5\textwidth]{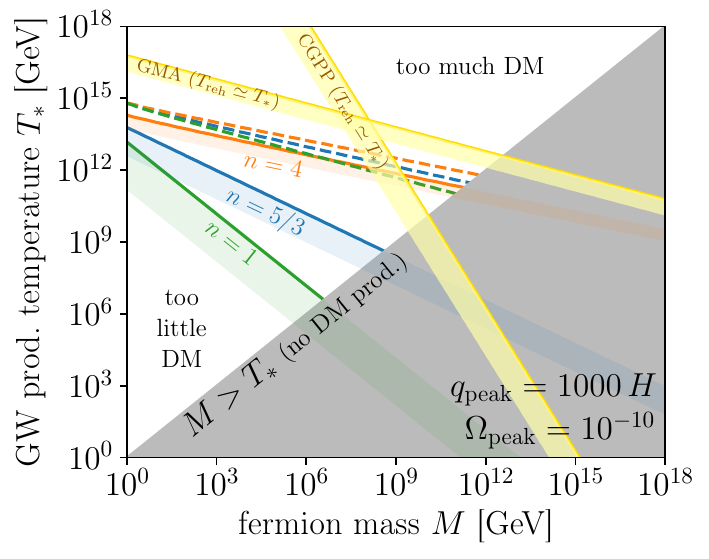}
    \end{tabular}
    \caption{Dark matter freeze-in from stochastic gravitational waves produced in a first-order phase transition. We have assumed a broken power law GW spectrum, with spectral index $m=3$ below the peak frequency, $q_\text{peak}$, and $n$ at higher frequencies (see \cref{Eq:Omega-PT}). Green, blue, and orange lines show the phase transition temperature $T_*$ and DM mass $M$ required to explain the observed DM density for different values of $n$. Solid and dashed lines correspond to incoherent and coherent GWs, respectively. Below the lines, GW-induced freeze-in can still contribute a fraction of the DM. The bottom edge of the shaded bands indicates where that fraction is 1\%. For comparison, we show in yellow the parameter regions in which conventional cosmological production of supermassive fermions by the expansion of the Universe \cite{Kolb:2017jvz, Ema:2019yrd, Kolb:2023ydq} and graviton-mediated inflaton annihilation \cite{Bernal:2018qlk, Clery:2021bwz} give the correct relic density. The different panels correspond to different GW peak amplitudes $\Omega_\text{peak}$ and peak frequencies $q_\text{peak}$, as indicated in the plots.}
    \label{fig:OmegaD}
\end{figure}

We plot our result in the $T_*$-vs-$M$ plane in \cref{fig:OmegaD}, demonstrating that our mechanism can explain the observed DM density in the Universe for a wide range of DM masses and temperature scales. As shown by the colored lines, the exact values of $T_*$ and $M$ for which the correct $\Omega_{\psi,0}$ is realized depend on the shape of the GW spectrum, but the mechanism typically favors $T_*$ well above the electroweak scale and safely below the Planck scale. The sensitivity to the shape of the GW spectrum is much larger for incoherent GW backgrounds (solid lines) than for coherent ones (dashed lines). In plotting \cref{fig:OmegaD}, we have chosen particular values of $\Omega_\text{peak}$ and $q_\text{peak}$ (with $\Omega_\text{peak}$ well below the Big Bang Nucleosynthesis bound, $\Omega_\text{peak} \lesssim \num{3.5e-6} \, m \, n / (m + n)$), but the scaling at different $\Omega_\text{peak}$, $q_\text{peak}$ can be read off immediately from \cref{Eq:OmegaD}.

Note that the values of $T_*$ and $q_\text{peak}$ for which our model explains the DM relic density correspond to GW peak frequencies today, $f_\text{peak}$, of order kHz to GHz.\footnote{To see this, note that
\begin{align}
    f_\text{peak} \equiv q_\text{peak}
        = \SI{4.8}{MHz} \times \bigg( \frac{q_\text{peak} / \mathcal{H}_*}{100} \bigg)
                               \bigg( \frac{g_*(\tau_*)}{106.75} \bigg)^{1/6}
                               \bigg( \frac{T_*}{\SI{3e11}{GeV}} \bigg),
\end{align}
as can be seen by invoking the definition of $\mathcal{H}_*$, \cref{eq:rad-era-relations}, the expression for the Hubble rate during radiation domination, $H(T) = \sqrt{8 \pi G / 3} \sqrt{\pi^2 g(T) / 30} \, T$, as well as entropy conservation, $g T^3 a^3 = \text{const}$.} The lower end of this range falls well within the sensitivity range of future terrestrial gravitational wave detectors like the Einstein Telescope \cite{Maggiore:2019uih} and Cosmic Explorer \cite{Evans:2021gyd}, which cover frequencies between $\sim \SI{1}{Hz}$ and \SI{10}{kHz}.  Sensitivity at higher frequencies requires novel detector technologies \cite{Aggarwal:2020olq}, though, in this early stage of development, none of them would be able to probe cosmological backgrounds. Pushing the model towards lower frequencies, possibly explaining the DM relic abundance with GW backgrounds accessible to LISA \cite{LISACosmologyWorkingGroup:2022jok} would require low values of $q_\text{peak} / \mathcal{H}_*$ (corresponding e.g.\ to very slow phase transitions), hard spectra (i.e.\ low $n$), and incoherent emission.

\subsection{Comparison with gravitational production of superheavy fermions}
\label{Sec:conventional-production}

Let us now compare this new production mechanism for (effectively) massless fermions by GWs with i) the conventional production mechanism of superheavy fermions by the expansion of the Universe, i.e.\ cosmological gravitational particle production (CGPP) \cite{Kolb:2017jvz, Ema:2019yrd}, and ii) dark matter production via $s$-channel graviton exchange in the hot primordial plasma ($S S \to h_{ij} \to X X$, where $S$ is a SM field or the inflaton) \cite{Bernal:2018qlk, Clery:2021bwz}. For an excellent review article on this topic see Ref.~\cite{Kolb:2023ydq}.

Consider superheavy fermions with a mass $M$ that satisfies $M < H_* < H_\text{inf}$, where $H_\text{inf}$ is the Hubble parameter at the end of inflation. The energy density of the dark fermions today through CGPP is \cite{Kolb:2023ydq}
\begin{align}
    \frac{\rho_{\Psi,0}^\text{CGPP}}{\rho_{\text{DM},0}}
        \sim 7 \times \bigg(\frac{M}{10^{11} \text{GeV}}\bigg)^2
                      \bigg( \frac{T_\text{reh}}{10^9 \text{GeV}}\bigg) ,
    \label{Eq:rho-cgpp}
\end{align}
with the reheating temperature $T_\text{reh}$. Note that this process is not thermal and the dependence on $T_\text{reh}$ enters through the redshift.  Here, we have used \cref{Eq:cH-App} for $\mathcal{H}_*$, assuming radiation domination, i.e.\ $T_* \lesssim T_\text{reh}$.

For high reheating temperatures, $T_{\text{reh}} \gtrsim \SI{e13}{GeV}$, the annihilation of inflatons or SM particles through $s$-channel graviton exchange presents an alternative gravitational production mechanism for dark matter \cite{Bernal:2018qlk, Clery:2021bwz}. The relic density of dark matter produced by graviton-mediated annihilation of inflatons (GMA) is \cite{Bernal:2018qlk, Kolb:2023ydq}
\begin{align}
    \frac{\rho_{\Psi,0}^\text{GMA}}{\rho_{\text{DM},0}}
        \sim  5 \times \bigg(\frac{M}{10^{12} \text{GeV}}\bigg)
                      \bigg( \frac{T_\text{reh}}{10^{13} \text{GeV}}\bigg)^3 ,
    \label{Eq:Omega-gma}
\end{align}
where $M \lesssim T_\text{reh}$ is required.

In \cref{fig:OmegaD}, the parameter regions in which cosmological gravitational particle production (\cref{Eq:rho-cgpp}) and graviton-mediated inflaton annihilation (\cref{Eq:Omega-gma}) yield between 1\% and 100\% of DM relic abundance are shown as yellow bands. We see that these mechanisms, especially CGPP, require larger $M$ and $T_*$ than gravitational-wave induced production. In other words, over vast regions of parameter, our loop-induced mechanism dominates over these tree-level mechanisms.

\section{Summary and Outlook}
\label{Sec:conclusion}

Our study in \cite{Maleknejad:2024ybn} has unveiled a new mechanism for the production of Weyl fermions in the early Universe: a gravitational freeze-in process of fermions driven by stochastic gravitational waves in an expanding Universe, i.e.\ GW-induced freeze-in. This is notable since the expansion of the Universe alone cannot changes the Weyl fermion number density. But at the loop level by considering cosmic perturbations, it does. Cosmological correlators are usually generated at the tree level with subleading loop contributions. In contrast, GW-induced freeze-in starts at the loop level. In this work, we have presented an extended study of this new phenomenon.  The crucial point is that a GW background introduces new scales in the system and breaks the fermions' conformal invariance. Our fermions are coupled to the rest of the SM only through minimal gravitational interactions, yet a sizeable relic abundance can be generated. As the fermions are weakly coupled and never reach thermal equilibrium, this abundance will survive until today. In \cref{tab:context} we have compared the new gravitational fermion production mechanism introduced in \cite{Maleknejad:2024ybn} to the other gravitational production mechanisms known in the literature.

We have undertaken a detailed study of GW-induced freeze-in using the in--in formalism during the radiation era. In particular, we have computed the 1-loop contribution of stochastic gravitational waves to the energy density (\cref{Eq:loop} and \cref{Eq:loop2}) of Weyl fermions. We have also calculated their pressure (\cref{eq:P}), showing that they indeed behave like radiation.  We have first obtained a general expression (\cref{Eq:rho-tot}), and have then evaluated it for specific gravitational wave backgrounds, notably those whose frequency spectrum can be parameterized as a broken power law (\cref{Eq:Omega-PT}). This includes in particular gravitational waves generated by phase transitions and primordial magnetic fields, and approximately also gauge fields in inflation and gauge reheating/preheating. Given that our result is related to the unequal-time 2-point function of GWs, we considered two extreme scenarios for the temporal coherence of GWs, namely fully incoherent in time and fully coherent in time. The energy density of Weyl fermions for the incoherent and coherent cases are given in \cref{Eq:rho-tot-incoh-2} and \cref{Eq:rho-coh-} respectively. 

The final Weyl fermion energy density is tied to the characteristics of the gravitational wave spectrum that sources it through its proportionality to \textit{i)} the peak wave number $q_\text{peak}$, \textit{ii)} the peak energy density $\Omega_\text{peak}$, as well as \textit{iii)} the degree of coherence in time of GWs and \textit{iv)} the shape of the spectrum through $\mathcal{C}_\text{incoh}$ and $\mathcal{C}_\text{coh}$. Additionally, the relic density today depends on \textit{v)} the redshift and the Hubble rate at fermion production, which is written in terms of temperature at phase transition $T_*$. This is why, despite being a non-thermal production process, our final result depends on $T_*$. 

If the initially massless fermions acquire a mass later in cosmological history, they can play the role of dark matter (see \cref{Eq:OmegaD,fig:OmegaD}). The typical scale $T_*$ of DM production in this scenario is between \SI{e6}{GeV} and \SI{e15}{GeV}. This corresponds to GW peak frequencies, $f_\text{peak}$, today ranging from kHz to GHz. The lower end of this range aligns with the sensitivity band of forthcoming terrestrial gravitational wave detectors such as the Einstein Telescope \cite{Maggiore:2019uih} and Cosmic Explorer \cite{Evans:2021gyd}, while higher frequencies present an interesting target for novel detection techniques currently under development \cite{Aggarwal:2020olq}. Of particular interest is the fact that our loop-induced gravitational fermion production mechanism works in regions of paramater space not accessible to tree-level gravitational production mechanisms, namely cosmological gravitational particle production (CGPP) and Graviton-Mediated-Annihilation (GMA) (\cref{Sec:conventional-production,fig:OmegaD}).

The next step in advancing this line of research is to go beyond our analytical estimates by means of numerical simulations to improve the precision of our predictions. In a separate work, we plan to apply this approach specifically to gravitational waves  generated by primordial magnetic fields \cite{in-preparation-I} and for fluctuations during inflation \cite{in-preparation-II}. Another possible direction of future research is the inclusion of CP violation through chiral gravitational waves, which could lead to a difference in the amounts of particles and anti-particles produced.

\section*{Acknowledgments}

We are grateful to Nima Arkani-Hamed, Cliff Burgess, Chiara Caprini, Xingang Chen, Tim Cohen, Raphael Flauger, Dan Green, Rocky Kolb, Eiichiro Komatsu, Andrew Long, Juan Maldacena, Enrico Pajer, Marco Simonovich, and Sasha Zhibeodov for very useful discussions.  We are thankful to Tim Cohen for his valuable feedback on the draft. AM is grateful to Hengameh Bagherian and Alberto Roper Pol for their collaborations on related topics. AM would like to thank CERN for hospitality during the final stages of this work. AM's work is supported by the Royal Society University Research Fellowship, Grant No. RE22432.

\appendix 
\section{Notations and Conventions}
\label{Sec:notation}

In this appendix, we summarize our notations and conventions. Throughout our work, a dot denotes a derivative with respect to cosmic time,
\begin{align}
    \dot X\equiv \p_t X,
\end{align}
and a prime denotes a derivative with respect to conformal time
\begin{align}
    X' \equiv \p_{\tau} X.   
\end{align}
We set the cosmic scale factor today to $a_0 = 1$, and we use the notation $\mpl = (8\pi G)^{-\frac12}$ for the reduced Planck mass.

The Hubble parameter $H$, conformal Hubble parameter $\mathcal{H}$, and conformal time $\tau$ during the radiation era are related as 
\begin{align}
    \mathcal{H} = a H = \frac{1}{\tau}.    
\end{align}
Throughout, the following notation is used for symmetric tensor products
\begin{align}
    X_{(i}Y_{j)} \equiv X_i Y_j + X_j Y_i.
\end{align}
Our convention for Fourier expansions follows ref.~\cite{Weinberg:2008zzc}, i.e.,
\begin{align}
    f(\bx,\tau) = \int \! d^3q \, f_q(\tau) \, e^{i \bq.\bx}.  
\end{align}
The following subscripts are frequently used: for a given quantity $X$,
\begin{align}
    \begin{split}
    X_i \quad \text{denotes} \quad & \text{the initial value}, \\
    X_* \quad \text{denotes} \quad & \text{the value at the settling time $t_*$ (see \cref{Sec:GW-spectrum})} \\
    X_0 \quad \text{denotes} \quad & \text{the value today}, \\
    X_I \quad \text{denotes} \quad & \text{an operator evaluated in the interaction picture}.
    \end{split}
\intertext{The following conventions are used for indices:}
    \begin{split}
    (\mu, \nu, \xi, \dots) \quad \text{denote} \quad
        & \text{space-time Lorentz indices}, \\
    (\alpha, \beta, \gamma, \dots) \quad  \text{denote} \quad
        & \text{tangent space coordinates}, \\
    (i, j, k, \dots) \quad \text{denote} \quad
        & \text{spatial Lorentz indices}, \\
    (a, b, c,\dots) \quad  \text{denote} \quad
        & \text{spatial tangent space coordinates}.
    \end{split}
\end{align}
The 4-component Dirac fermions are denoted as $\boldsymbol{\Psi}_{\!D}=(\boldsymbol{\Psi}_{L},\boldsymbol{\Psi}_R)$ where $\boldsymbol{\Psi}_{L,R}$ are Weyl fermions.  In this manuscript, we mostly work with 2-component spinors which are acted upon by $2 \times 2$ matrices, and we use bold symbols for both. In particular, $\boldsymbol{\uppsi}$ is a 2-spinor and
\begin{align}
    \boldsymbol{\Psi} \equiv a^{3/2} \boldsymbol{\uppsi},    
\end{align}
is the corresponding canonically normalized spinor. We have two complete sets of unitary $2 \times 2$ matrices,
\begin{align}
    \bs^{\alpha}       \equiv ( \mathbf{I}_2,  \bs^i) \an
    \bar{\bs}^{\alpha} \equiv ( \mathbf{I}_2, -\bs^i),
\end{align}
in which $\bs^i$ are the Pauli matrices. In addition, we use boldface notation also for the comoving 3-vectors $\bx$, $\bk$, $\bq$. We denote tetrads, GW helicity states, and spinors as
\begin{align}
    {\bf{e}}^\alpha_\mu  &= ~\text{tetrads},\\
    \E^s_{ij}            &= ~\text{GW helicity states}
                             (s=\pm 2),\\
    {\bf{E}}^\lambda_\bk &= ~\text{spinor helicity states}
                             (\lambda = \pm \tfrac12),
\end{align}
where $s$ and $\lambda$ label the spins of GWs and fermions, respectively.

\section{Spinors in Curved Space}
\label{Sec:app-spin-connection}

Here,  we work out the spin connections of the FLRW metric with gravitational waves up to second order in GW perturbation theory. We fix the gauge to second order in small fluctuations. The perturbed metric can be written as
\begin{align}
    g_{00} = -1  \an  g_{ij} = a^2 \hat{g}_{ij},    
    \label{Eq:pert-FLRW}
\end{align}
with
\begin{align}
    \det\hat{g}  = 1, \quad
    \hat{g}_{ij} = (\delta_{ij} + h_{ij} + \tfrac12 h_{ik} h_{jk} + \dots).
\end{align}
We have $h_{ii} = 0$ and $\p_i h_{ij} = 0$, up to second order in perturbation theory.  The inverse metric is given by
\begin{align}
    g^{ij} = \frac{1}{a^2} (\delta_{ij} - h_{ij} + \tfrac12 h_{ik} h_{jk} + \dots).
\end{align}
The metric can be expressed in terms of the tetrads $\{ \e^\alpha_{~\mu} \}$,
\begin{align}
    g_{\mu\nu} = \e^\alpha_{~\mu} \e^\beta_{~\nu} \eta_{\alpha\beta},    
\end{align}
and from \cref{Eq:pert-FLRW} it is then straightforward to show that the non-zero components of the tetrads are
\begin{align}
    \e_{~0}^0       &= 1, \\ 
    {\bf{e}}_{~i}^a &= a \, \delta^a_{j} (\delta_{ij} + \tfrac12 h_{ij}
                                        + \tfrac18 h_{ik} h_{jk}),
\end{align}
The metric connections (Christoffel symbols) are
\begin{align}
    \Gamma^0_{~ij} &= a^2 \Big[ H \Big( \delta_{ij} + h_{ij} + \frac12 h_{ik}h_{jk} \Big)
                              + \frac12 \p_0 h_{ij} + \frac14 \p_0 (h_{ik} h_{jk}) \Big]  ,\\
    \Gamma^i_{~0j} &= H \delta_{ij} + \frac12 \p_0 h_{ij} + \frac14 \p_0(h_{ik} h_{jk})
                    - \frac12 h_{ik} \p_0 h_{jk}, \\
    \Gamma^i_{~jk} &= \frac12 \Big( -\p_i h_{jk} + \p_j h_{ik} + \p_k h_{ij} \Big)
                    + \frac14  \Big[ - \p_i (h_{jl} h_{kl}) + \p_j (h_{il} h_{kl})
                                     + \p_k (h_{il} h_{jl}) \Big] \nonumber\\
                   &\hspace{7cm}
                    + \frac12 \big( h_{il} \p_l h_{jk} - h_{il} \p_j h_{lk}
                                  - h_{il} \p_k h_{jl} \big).    
\end{align}
The spinor covariant derivative $\mathcal{D}_\mu$ is
\begin{align}
    \mathcal{D}_\mu \equiv \p_\mu - \omega_\mu,
    \label{Eq:slashD}
\end{align}
where $\omega_\mu \equiv \frac{i}{2} \omega^{\alpha\beta}_{~~\mu} \Sigma_{\alpha\beta}$ is the spin connection, 
$\Sigma_{\alpha\beta} = \frac{i}{4} [\gamma_{\alpha}, \gamma_{\beta}]$ are the generators of the Lorentz group in the spinor representation for Dirac fields, and
\begin{align}
    {\bf{\omega}}^{\alpha\beta}_{~~\mu}
         \equiv {\bf{e}}^{\alpha\nu} \nabla_{\mu} {\bf{e}}^{\beta}_{~\nu}
         = {\bf{e}}^{\alpha\nu} \p_{\mu} {\bf{e}}^{\beta}_{~\nu}
         - \Gamma^\nu_{\mu\xi} {\bf{e}}^{\alpha\xi} {\bf{e}}^\beta_{~\nu}.
\end{align}
Here we work out the spin connections for a Dirac field and it is straightforward to find it for the left- and right-handed Weyl fermions. Expanding around the background up to first order in the GW amplitude yields
\begin{align}
    \delta_1\omega^{0a}_{~~0} &= \delta_1\omega^{ab}_{~~0} = 0, \\
    \delta_1\omega^{0a}_{~~j} &= a \Big[ H \delta^a_j
        + \frac12 \delta^{ak} \Big( H h_{jk} + \p_0 h_{jk} \Big) \Big],\\
    \delta_1\omega^{ab}_{~~i} &= \frac12 \delta^{al} \delta^{bk}
                         \Big[ \p_k h_{il} - \p_l h_{ik} \Big].    
\end{align}
At second order, we have
\begin{align}
    \delta_2\omega^{0a}_{~~0} &= 0, \\
    \delta_2\omega^{ab}_{~~0} &= \frac18 \delta_{j}^a \delta^b_i 
                                 \big( h_{il}\p_0 h_{jl} - h_{jl}\p_0 h_{il} \big) , \\ 
    \delta_2\omega^{0a}_{~~j} &= a \delta^a_k \Big( \frac18 H h_{jl} h_{kl}
                                                  + \frac14 h_{lj}\p_0 h_{kl} \Big) ,\\
    \delta_2\omega^{ab}_{~~i} &= \frac14 \delta^a_j \delta^b_k \Big(
        \frac12 h_{kl} \p_ih_{jl} - \frac12 h_{jl} \p_ih_{kl}  + h_{il} \p_k h_{jl}
      - h_{il} \p_j h_{kl} + h_{jl} \p_l h_{ik} - h_{kl} \p_l h_{ij} \Big),
\end{align}
where $\delta_2$ denotes second order in $h_{ij}$.  The full spin connections up to 2nd order in $h_{ij}$ are
\begin{align}
    \omega_0 & = \frac{i}{16} \epsilon^{kja} h_{ij} \p_0 h_{ik}   
                \begin{pmatrix}
                    {\boldsymbol{\sigma}}_a & 0 \\
                    0                       & {\boldsymbol{\sigma}}_a
                \end{pmatrix}  ,\\
    \omega_i     &= \frac{a}{2} \bigg[ H \delta^a_i
              + \frac12 \delta^{ak} \bigg(
                    H h_{ik} + \p_0 h_{ik}
                  + \frac12 h_{li} \Big( \p_0 h_{kl}
                                       + \frac{H}{2} h_{kl} \Big)
                \bigg) \bigg]
                \begin{pmatrix}
                    {\boldsymbol{\sigma}}_a & 0 \\
                    0                       & -{\boldsymbol{\sigma}}_a
                \end{pmatrix} \nonumber\\
             &
              + \frac{i}{2} \epsilon_{lk}^{~~a} \bigg[
                    \p_k h_{il} + \frac12 \Big(
                        \frac12 h_{kj} \p_i h_{lj}
                      + h_{ij} \p_k h_{lj}  + h_{lj} \p_j h_{ik} 
                    \Big)
                \bigg]
                \begin{pmatrix}
                    {\boldsymbol{\sigma}}_a & 0 \\
                    0                       & {\boldsymbol{\sigma}}_a
                \end{pmatrix}.
\end{align}

\subsection{First order Dirac operator}
\label{Sec:1st-GW-fermion}

We now work out the Dirac equation in the presence of GWs. The general form of the Dirac equation for a massless field in curved space is
\begin{align}
   {\bf{e}}^{~\mu}_\alpha \gamma^\alpha \mathcal{D}_\mu \boldsymbol{\uppsi} = 0.
   \label{Eq:Dirac}
\end{align}
Up to first order in $h_{ij}$, we have
\begin{align}
   & {\bf{e}}^{~\mu}_\alpha \gamma^\alpha \mathcal{D}_\mu =\nonumber\\
        &= \gamma^0 \p_{0} + \gamma^b {\bf{e}}^{~i}_b \bigg(
               \p_i - \frac{a}{2} \bigg[
                   H \delta^a_i
                 + \frac12 \delta^{ak} \big( H h_{ik} + \p_0 h_{ik} \big)
               \bigg] \begin{pmatrix}
                          {\boldsymbol\sigma}_a & 0 \\ 
                          0                  & -{\boldsymbol\sigma}_a
                      \end{pmatrix}
       + \frac{i}{4} \epsilon_{lk}^{~~a} \p_l h_{ik}
           \begin{pmatrix} 
               {\boldsymbol\sigma}_a & 0 \\
               0                     & {\boldsymbol\sigma}_a
           \end{pmatrix} \bigg) \nonumber\\
%
%
        &= \gamma^0 \p_{0} + \bigg\{ \gamma^i \frac1a (\p_i - \frac12 h_{ij} \p_j) 
         - \frac{1}{2} \bigg[ H \gamma^k
               + \frac12 \gamma^i \big( H h_{ik} + \p_0 h_{ik} \big)
           \bigg] \begin{pmatrix}
                      {\boldsymbol\sigma}_k & 0 \\
                      0                     & -{\boldsymbol\sigma}_k
                  \end{pmatrix} \nonumber\\
       & - \frac{1}{4a} \big( \gamma^j \p_i h_{ij}
                            - \gamma^j \p_j h_{ii} \big) \bigg\} . \,
\end{align}
Imposing the traceless and transverse condition on $h_{ij}$, we find the first-order Dirac operator as
\begin{align}\label{Eq:rescale}
    {\bf{e}}^{~\mu}_{\alpha} \gamma^{\alpha}\mathcal{D}_{\mu}
        &= \gamma^0 \Big( \p_{0} + \frac32 H \Big)
         + \frac1a \gamma^i \Big(\p_i - \frac12 h_{ij}\p_j \Big)
\end{align}
This implies that the canonically renormalized Dirac fermion field is $\Psi_D \equiv a^{\frac32}\uppsi_D$. The first-order interaction Lagrangian can be simplified as
\begin{align}
   \delta_1 \mathcal{L}_\text{int} = -\frac{i}{2a^4} h_{ij}\bar{\boldsymbol{\Psi}}_{\!D} \gamma^{i} \overset{\leftrightarrow}{\p}_j \boldsymbol{\Psi}_{\!D}.
\end{align}
The field equation in terms of $\Psi$ is
\begin{align}
    \Big( \gamma^0 \p_0 + \frac{1}{a} \gamma^i \p_i
        - \frac12 \frac{h_{ij}}{a} \gamma^j \p_i \Big) \Psi_{\!D} =0.    
\end{align}
As we see, the effect of the expansion of the Universe has disappeared in the dynamics of the canonically normalized field. This is a consequence of the conformal symmetry of the Weyl fermions.

\subsection{Second order Dirac operator}
\label{Sec:2nd-GW-fermion}

At second order in $h_{ij}$,  we have 4 contributions, namely
\begin{align}
    \delta_2(\gamma^0 \omega_0) &=
        \frac{i}{16} \epsilon^{kja} h_{ij} \p_0 h_{ik}
        \begin{pmatrix}
            0                       & {\boldsymbol\sigma}_a \\ 
            {\boldsymbol{\sigma}}_a & 0
        \end{pmatrix} , \\ 
    \gamma^b \delta_1 {\bf{e}}^{~i}_b \delta_1\omega_i &=  
        \frac18 \gamma^b h_{ij} \delta_b^j \delta^{ak} \bigg[
            \big( H h_{ik} + \p_0 h_{ik} \big) \begin{pmatrix}
                                                   {\boldsymbol\sigma}_a & 0 \\
                                                   0                     & -{\boldsymbol\sigma}_a
                                                \end{pmatrix}
          + \frac{i}{a} \epsilon_{ln}^{~~k} \p_l h_{in}
            \begin{pmatrix}
                {\boldsymbol\sigma}_a & 0 \\
                0                     & {\boldsymbol\sigma}_a
            \end{pmatrix}
        \bigg] , \nonumber \\
    \gamma^b \delta_2{\bf{e}}^{~i}_b \omega_i  &=  
        -\frac{H}{16} \gamma^b \delta_b^j \delta^{ak} h_{ij} h_{ik}
        \begin{pmatrix}
            {\boldsymbol{\sigma}}_a & 0 \\
            0                       & -{\boldsymbol\sigma}_a
        \end{pmatrix}, \nonumber \\
    \gamma^b {\bf{e}}^{~i}_b \delta_2\omega_i  &=
        - \frac{i}{8a} \gamma^b \delta^j_b \delta^{ak} \epsilon_{ln}^{~~k} \bigg[
            - \frac12 h_{in} \p_j h_{il}  +  h_{ij} \p_l h_{in}  +  h_{in} \p_i h_{jl}
        \bigg] \begin{pmatrix} 
                   {\boldsymbol\sigma}_a & 0 \\
                   0                     & {\boldsymbol\sigma}_a
                \end{pmatrix} \nonumber \\
    &\quad
        -\frac{1}{8} \gamma^b h_{ij} \delta^j_b \delta^{ak}
        \Big( \p_0 h_{ik} + \frac{H}{2} h_{i k} \Big)
        \begin{pmatrix}
            {\boldsymbol\sigma}_a & 0 \\
            0                     & -{\boldsymbol\sigma}_a
        \end{pmatrix}.
\end{align}
These terms can be combined into 
\begin{align}
    \delta_2({\bf{e}}^{~\mu}_{\alpha} \gamma^{\alpha}\mathcal{D}_{\mu}) &= 
        \gamma^0 \frac{i}{16a} \epsilon_{ln}^{~~j} h_{in}\p_j h_{il} \gamma^5
      + \frac1a \gamma^i \Big( \frac{1}{16} h_{kj}\p_k h_{ij}
                   - \frac{i a}{16} \epsilon^{~~j}_{in} h_{lj} \p_0 h_{ln} \gamma^5 \Big) \nonumber\\
         &= 
         \frac{i}{16} \big( \frac1a \gamma^0\epsilon_{ln}^{~~j} h_{in}\p_j h_{il} 
                         - \epsilon^{~~j}_{ln} h_{ij} \p_0 h_{in} \gamma^l \big) \gamma^5,      
\end{align}
where in the second line we have dropped a total derivative.  The second-order interaction Lagrangian can therefore be simplified as
\begin{align}
   \delta_2 \mathcal{L}_\text{int} = -\frac{i}{16a^3} e^{\mu}_{~\alpha} h_{ij} \p_{\mu} h_{ik}\bar{\boldsymbol{\Psi}}_{\!D} \Gamma^{\alpha j k} \boldsymbol{\Psi}_{\!D},
\end{align}
where $\Gamma^{\alpha jk}$ is the totally anti-symmetrized product of three gamma matrices.

\bibliographystyle{JHEP}
\bibliography{ref}

\providecommand{\href}[2]{#2}\begingroup\raggedright\begin{thebibliography}{10}

\bibitem{Maleknejad:2024ybn}
A.~Maleknejad and J.~Kopp, {\it {Gravitational Wave-Induced Freeze-In of Fermionic Dark Matter}},  \href{http://arxiv.org/abs/2405.09723}{{\tt arXiv:2405.09723}}.

\bibitem{Sorbo:2011rz}
L.~Sorbo, {\it {Parity violation in the Cosmic Microwave Background from a pseudoscalar inflaton}},  {\em JCAP} {\bf 06} (2011) 003, [\href{http://arxiv.org/abs/1101.1525}{{\tt arXiv:1101.1525}}].

\bibitem{Maleknejad:2016qjz}
A.~Maleknejad, {\it {Axion Inflation with an SU(2) Gauge Field: Detectable Chiral Gravity Waves}},  {\em JHEP} {\bf 07} (2016) 104, [\href{http://arxiv.org/abs/1604.03327}{{\tt arXiv:1604.03327}}].

\bibitem{Komatsu:2022nvu}
E.~Komatsu, {\it {New physics from the polarized light of the cosmic microwave background}},  {\em Nature Rev. Phys.} {\bf 4} (2022), no.~7 452--469, [\href{http://arxiv.org/abs/2202.13919}{{\tt arXiv:2202.13919}}].

\bibitem{Witten:1984rs}
E.~Witten, {\it {Cosmic Separation of Phases}},  {\em Phys. Rev. D} {\bf 30} (1984) 272--285.

\bibitem{Schwaller:2015tja}
P.~Schwaller, {\it {Gravitational Waves from a Dark Phase Transition}},  {\em Phys. Rev. Lett.} {\bf 115} (2015), no.~18 181101, [\href{http://arxiv.org/abs/1504.07263}{{\tt arXiv:1504.07263}}].

\bibitem{Caprini:2015zlo}
C.~Caprini et~al., {\it {Science with the space-based interferometer eLISA. II: Gravitational waves from cosmological phase transitions}},  {\em JCAP} {\bf 04} (2016) 001, [\href{http://arxiv.org/abs/1512.06239}{{\tt arXiv:1512.06239}}].

\bibitem{RoperPol:2023bqa}
A.~Roper~Pol, A.~Neronov, C.~Caprini, T.~Boyer, and D.~Semikoz, {\it {LISA and $\gamma$-ray telescopes as multi-messenger probes of a first-order cosmological phase transition}},  \href{http://arxiv.org/abs/2307.10744}{{\tt arXiv:2307.10744}}.

\bibitem{Brandenburg:2021aln}
A.~Brandenburg, Y.~He, T.~Kahniashvili, M.~Rheinhardt, and J.~Schober, {\it {Relic gravitational waves from the chiral magnetic effect}},  {\em Astrophys. J.} {\bf 911} (2021), no.~2 110, [\href{http://arxiv.org/abs/2101.08178}{{\tt arXiv:2101.08178}}].

\bibitem{RoperPol:2021xnd}
A.~Roper~Pol, S.~Mandal, A.~Brandenburg, and T.~Kahniashvili, {\it {Polarization of gravitational waves from helical MHD turbulent sources}},  {\em JCAP} {\bf 04} (2022), no.~04 019, [\href{http://arxiv.org/abs/2107.05356}{{\tt arXiv:2107.05356}}].

\bibitem{Adshead:2019igv}
P.~Adshead, J.~T. Giblin, M.~Pieroni, and Z.~J. Weiner, {\it {Constraining Axion Inflation with Gravitational Waves across 29 Decades in Frequency}},  {\em Phys. Rev. Lett.} {\bf 124} (2020), no.~17 171301, [\href{http://arxiv.org/abs/1909.12843}{{\tt arXiv:1909.12843}}].

\bibitem{Figueroa:2022iho}
D.~G. Figueroa, A.~Florio, N.~Loayza, and M.~Pieroni, {\it {Spectroscopy of particle couplings with gravitational waves}},  {\em Phys. Rev. D} {\bf 106} (2022), no.~6 063522, [\href{http://arxiv.org/abs/2202.05805}{{\tt arXiv:2202.05805}}].

\bibitem{Hindmarsh:1994re}
M.~B. Hindmarsh and T.~W.~B. Kibble, {\it {Cosmic strings}},  {\em Rept. Prog. Phys.} {\bf 58} (1995) 477--562, [\href{http://arxiv.org/abs/hep-ph/9411342}{{\tt hep-ph/9411342}}].

\bibitem{Auclair:2019wcv}
P.~Auclair et~al., {\it {Probing the gravitational wave background from cosmic strings with LISA}},  {\em JCAP} {\bf 04} (2020) 034, [\href{http://arxiv.org/abs/1909.00819}{{\tt arXiv:1909.00819}}].

\bibitem{Ford:1986sy}
L.~H. Ford, {\it {Gravitational Particle Creation and Inflation}},  {\em Phys. Rev. D} {\bf 35} (1987) 2955.

\bibitem{Chung:1998zb}
D.~J.~H. Chung, E.~W. Kolb, and A.~Riotto, {\it {Superheavy dark matter}},  {\em Phys. Rev. D} {\bf 59} (1998) 023501, [\href{http://arxiv.org/abs/hep-ph/9802238}{{\tt hep-ph/9802238}}].

\bibitem{Parker_Toms_2009}
L.~Parker and D.~Toms, {\em Quantum Field Theory in Curved Spacetime: Quantized Fields and Gravity}.
\newblock Cambridge Monographs on Mathematical Physics. Cambridge University Press, 2009.

\bibitem{Kolb:2023ydq}
E.~W. Kolb and A.~J. Long, {\it {Cosmological gravitational particle production and its implications for cosmological relics}},  \href{http://arxiv.org/abs/2312.09042}{{\tt arXiv:2312.09042}}.

\bibitem{Kolb:2017jvz}
E.~W. Kolb and A.~J. Long, {\it {Superheavy dark matter through Higgs portal operators}},  {\em Phys. Rev. D} {\bf 96} (2017), no.~10 103540, [\href{http://arxiv.org/abs/1708.04293}{{\tt arXiv:1708.04293}}].

\bibitem{Ema:2019yrd}
Y.~Ema, K.~Nakayama, and Y.~Tang, {\it {Production of purely gravitational dark matter: the case of fermion and vector boson}},  {\em JHEP} {\bf 07} (2019) 060, [\href{http://arxiv.org/abs/1903.10973}{{\tt arXiv:1903.10973}}].

\bibitem{Garny:2015sjg}
M.~Garny, M.~Sandora, and M.~S. Sloth, {\it {Planckian Interacting Massive Particles as Dark Matter}},  {\em Phys. Rev. Lett.} {\bf 116} (2016), no.~10 101302, [\href{http://arxiv.org/abs/1511.03278}{{\tt arXiv:1511.03278}}].

\bibitem{Bernal:2018qlk}
N.~Bernal, M.~Dutra, Y.~Mambrini, K.~Olive, M.~Peloso, and M.~Pierre, {\it {Spin-2 Portal Dark Matter}},  {\em Phys. Rev. D} {\bf 97} (2018), no.~11 115020, [\href{http://arxiv.org/abs/1803.01866}{{\tt arXiv:1803.01866}}].

\bibitem{Clery:2021bwz}
S.~Clery, Y.~Mambrini, K.~A. Olive, and S.~Verner, {\it {Gravitational portals in the early Universe}},  {\em Phys. Rev. D} {\bf 105} (2022), no.~7 075005, [\href{http://arxiv.org/abs/2112.15214}{{\tt arXiv:2112.15214}}].

\bibitem{Greene:1998nh}
P.~B. Greene and L.~Kofman, {\it {Preheating of fermions}},  {\em Phys. Lett. B} {\bf 448} (1999) 6--12, [\href{http://arxiv.org/abs/hep-ph/9807339}{{\tt hep-ph/9807339}}].

\bibitem{Adshead:2018oaa}
P.~Adshead, L.~Pearce, M.~Peloso, M.~A. Roberts, and L.~Sorbo, {\it {Phenomenology of fermion production during axion inflation}},  {\em JCAP} {\bf 06} (2018) 020, [\href{http://arxiv.org/abs/1803.04501}{{\tt arXiv:1803.04501}}].

\bibitem{Maleknejad:2019hdr}
A.~Maleknejad, {\it {Dark Fermions and Spontaneous $CP$ violation in $SU(2)$-axion Inflation}},  {\em JHEP} {\bf 07} (2020) 154, [\href{http://arxiv.org/abs/1909.11545}{{\tt arXiv:1909.11545}}].

\bibitem{Maleknejad:2020yys}
A.~Maleknejad, {\it {SU(2)R and its axion in cosmology: A common origin for inflation, cold sterile neutrinos, and baryogenesis}},  {\em Phys. Rev. D} {\bf 104} (2021), no.~8 083518, [\href{http://arxiv.org/abs/2012.11516}{{\tt arXiv:2012.11516}}].

\bibitem{Maleknejad:2020pec}
A.~Maleknejad, {\it {Chiral anomaly in SU(2)$_{R}$-axion inflation and the new prediction for particle cosmology}},  {\em JHEP} {\bf 21} (2020) 113, [\href{http://arxiv.org/abs/2103.14611}{{\tt arXiv:2103.14611}}].

\bibitem{Zhang:2023xcd}
R.~Zhang, Z.~Xu, and S.~Zheng, {\it {Gravitational freeze-in dark matter from Higgs preheating}},  {\em JCAP} {\bf 07} (2023) 048, [\href{http://arxiv.org/abs/2305.02568}{{\tt arXiv:2305.02568}}].

\bibitem{Alexander:2004us}
S.~H.-S. Alexander, M.~E. Peskin, and M.~M. Sheikh-Jabbari, {\it {Leptogenesis from gravity waves in models of inflation}},  {\em Phys. Rev. Lett.} {\bf 96} (2006) 081301, [\href{http://arxiv.org/abs/hep-th/0403069}{{\tt hep-th/0403069}}].

\bibitem{Maleknejad:2016dci}
A.~Maleknejad, {\it {Gravitational leptogenesis in axion inflation with SU(2) gauge field}},  {\em JCAP} {\bf 12} (2016) 027, [\href{http://arxiv.org/abs/1604.06520}{{\tt arXiv:1604.06520}}].

\bibitem{Maleknejad:2014wsa}
A.~Maleknejad, {\it {Chiral Gravity Waves and Leptogenesis in Inflationary Models with non-Abelian Gauge Fields}},  {\em Phys. Rev. D} {\bf 90} (2014), no.~2 023542, [\href{http://arxiv.org/abs/1401.7628}{{\tt arXiv:1401.7628}}].

\bibitem{Adshead:2015jza}
P.~Adshead and E.~I. Sfakianakis, {\it {Leptogenesis from left-handed neutrino production during axion inflation}},  {\em Phys. Rev. Lett.} {\bf 116} (2016), no.~9 091301, [\href{http://arxiv.org/abs/1508.00881}{{\tt arXiv:1508.00881}}].

\bibitem{Caldwell:2017chz}
R.~R. Caldwell and C.~Devulder, {\it {Axion Gauge Field Inflation and Gravitational Leptogenesis: A Lower Bound on B Modes from the Matter-Antimatter Asymmetry of the Universe}},  {\em Phys. Rev. D} {\bf 97} (2018), no.~2 023532, [\href{http://arxiv.org/abs/1706.03765}{{\tt arXiv:1706.03765}}].

\bibitem{Alvarez-Gaume:1983ihn}
L.~Alvarez-Gaume and E.~Witten, {\it {Gravitational Anomalies}},  {\em Nucl. Phys. B} {\bf 234} (1984) 269.

\bibitem{Eguchi:1976db}
T.~Eguchi and P.~G.~O. Freund, {\it {Quantum Gravity and World Topology}},  {\em Phys. Rev. Lett.} {\bf 37} (1976) 1251.

\bibitem{Caprini:2009yp}
C.~Caprini, R.~Durrer, and G.~Servant, {\it {The stochastic gravitational wave background from turbulence and magnetic fields generated by a first-order phase transition}},  {\em JCAP} {\bf 12} (2009) 024, [\href{http://arxiv.org/abs/0909.0622}{{\tt arXiv:0909.0622}}].

\bibitem{Caprini:2018mtu}
C.~Caprini and D.~G. Figueroa, {\it {Cosmological Backgrounds of Gravitational Waves}},  {\em Class. Quant. Grav.} {\bf 35} (2018), no.~16 163001, [\href{http://arxiv.org/abs/1801.04268}{{\tt arXiv:1801.04268}}].

\bibitem{in-preparation-II}
H.~Bagherian and A.~Maleknejad, {\it {Weyl Fermions and Gravitational Waves in Inflation}},  \href{http://arxiv.org/abs/[2024, in preparation]}{{\tt [2024, in preparation]}}.

\bibitem{Weinberg:2008zzc}
S.~Weinberg, {\em {Cosmology}}.
\newblock 2008.

\bibitem{Maggiore:2018sht}
M.~Maggiore, {\em {Gravitational Waves. Vol. 2: Astrophysics and Cosmology}}.
\newblock Oxford University Press, 3, 2018.

\bibitem{Donoghue:2017pgk}
J.~F. Donoghue, M.~M. Ivanov, and A.~Shkerin, {\it {EPFL Lectures on General Relativity as a Quantum Field Theory}},  \href{http://arxiv.org/abs/1702.00319}{{\tt arXiv:1702.00319}}.

\bibitem{Ortin:2015hya}
T.~Ortin, {\em {Gravity and Strings}}.
\newblock Cambridge Monographs on Mathematical Physics. Cambridge University Press, 2nd ed.~ed., 7, 2015.

\bibitem{Weinberg:2005vy}
S.~Weinberg, {\it {Quantum contributions to cosmological correlations}},  {\em Phys. Rev. D} {\bf 72} (2005) 043514, [\href{http://arxiv.org/abs/hep-th/0506236}{{\tt hep-th/0506236}}].

\bibitem{Senatore:2012nq}
L.~Senatore and M.~Zaldarriaga, {\it {On Loops in Inflation II: IR Effects in Single Clock Inflation}},  {\em JHEP} {\bf 01} (2013) 109, [\href{http://arxiv.org/abs/1203.6354}{{\tt arXiv:1203.6354}}].

\bibitem{Maggiore:2007ulw}
M.~Maggiore, {\em {Gravitational Waves. Vol. 1: Theory and Experiments}}.
\newblock Oxford University Press, 2007.

\bibitem{Strominger:2017zoo}
A.~Strominger, {\em {Lectures on the Infrared Structure of Gravity and Gauge Theory}}.
\newblock 3, 2017.

\bibitem{Durrer:2010xc}
R.~Durrer, {\it {Gravitational waves from cosmological phase transitions}},  {\em J. Phys. Conf. Ser.} {\bf 222} (2010) 012021, [\href{http://arxiv.org/abs/1002.1389}{{\tt arXiv:1002.1389}}].

\bibitem{RoperPol:2022iel}
A.~Roper~Pol, C.~Caprini, A.~Neronov, and D.~Semikoz, {\it {Gravitational wave signal from primordial magnetic fields in the Pulsar Timing Array frequency band}},  {\em Phys. Rev. D} {\bf 105} (2022), no.~12 123502, [\href{http://arxiv.org/abs/2201.05630}{{\tt arXiv:2201.05630}}].

\bibitem{Caprini:2009fx}
C.~Caprini, R.~Durrer, T.~Konstandin, and G.~Servant, {\it {General Properties of the Gravitational Wave Spectrum from Phase Transitions}},  {\em Phys. Rev. D} {\bf 79} (2009) 083519, [\href{http://arxiv.org/abs/0901.1661}{{\tt arXiv:0901.1661}}].

\bibitem{Scully_Zubairy_1997}
M.~O. Scully and M.~S. Zubairy, {\em {Quantum Optics}}.
\newblock 1997.

\bibitem{Maggiore:2019uih}
M.~Maggiore et~al., {\it {Science Case for the Einstein Telescope}},  {\em JCAP} {\bf 03} (2020) 050, [\href{http://arxiv.org/abs/1912.02622}{{\tt arXiv:1912.02622}}].

\bibitem{Evans:2021gyd}
M.~Evans et~al., {\it {A Horizon Study for Cosmic Explorer: Science, Observatories, and Community}},  \href{http://arxiv.org/abs/2109.09882}{{\tt arXiv:2109.09882}}.

\bibitem{Aggarwal:2020olq}
N.~Aggarwal et~al., {\it {Challenges and opportunities of gravitational-wave searches at MHz to GHz frequencies}},  {\em Living Rev. Rel.} {\bf 24} (2021), no.~1 4, [\href{http://arxiv.org/abs/2011.12414}{{\tt arXiv:2011.12414}}].

\bibitem{LISACosmologyWorkingGroup:2022jok}
{\bf LISA Cosmology Working Group} Collaboration, P.~Auclair et~al., {\it {Cosmology with the Laser Interferometer Space Antenna}},  {\em Living Rev. Rel.} {\bf 26} (2023), no.~1 5, [\href{http://arxiv.org/abs/2204.05434}{{\tt arXiv:2204.05434}}].

\bibitem{in-preparation-I}
A.~Roper~Pol, A.~Maleknejad, and J.~Kopp, {\it {Simulating Fermion Production by Gravitational Waves Background}},  \href{http://arxiv.org/abs/[2024, in preparation]}{{\tt [2024, in preparation]}}.

\end{thebibliography}\endgroup

\end{document}